\documentclass[10pt,twocolumn,letterpaper]{article}

\usepackage{cvpr}
\usepackage{times}
\usepackage{graphicx}
\usepackage{amsmath}
\usepackage{diagbox}
\usepackage{slashbox}
\usepackage{amssymb}
\usepackage{color, flushend}
\usepackage{subcaption}
\usepackage[style=base]{caption}
\usepackage[breaklinks=true,bookmarks=false]{hyperref}

\cvprfinalcopy 

\begin{document}

\title{\emph{Colored Kimia Path24} Dataset:\\ Configurations and Benchmarks with Deep Embeddings}

\author{Sobhan Shafiei,  Morteza Babaie, Shivam Kalra, H.R.Tizhoosh \\
Kimia Lab, University of Waterloo, Canada\\
{\tt\small \url{https://kimia.uwaterloo.ca/}}
}

\maketitle

\begin{abstract}
 The Kimia Path24 dataset has been introduced as a classification and retrieval dataset for digital pathology. Although it provides multi-class data, the color information has been neglected in the process of extracting patches. The staining information plays a major role in the recognition of tissue patterns. To address this drawback, we introduce the color version of Kimia Path24 by recreating sample patches from all 24 scans to propose \emph{Kimia Path24C}. We run extensive experiments to determine the best configuration for selected patches. To provide preliminary results for setting a benchmark for the new dataset, we utilize \textit{VGG16}, \textit{InceptionV3} and \textit{DenseNet-121} model as feature extractors. Then, we use these feature vectors to retrieve test patches. The accuracy of image retrieval  using \textit{DenseNet} was 95.92\% while the highest accuracy using \textit{InceptionV3}  and \textit{VGG16} reached 92.45\% and 92\%, respectively. We also experimented with ``deep barcodes'' and established that with a small loss in accuracy (e.g., 93.43\% for binarized features for DenseNet instead of 95.92\% when the features themselves are used), the search operations can be significantly accelerated. 
\end{abstract}

\section{Introduction}
Diagnosis of biopsy samples is performed by a pathologist examining the stained specimen on the glass slide using a microscope. In recent years, attempts have been made to capture the entire slide with a scanner and save it as a digital image or a whole slide image (WSI) \cite{pantanowitz2010digital}.
Digital pathology opens new horizons for biomedical research and clinical routine but also creates several challenges for computer vision research \cite{mccann2015automated,tizhoosh2018artificial}. Laboratories can scan a large number of slides per day, each being a gigapixel image containing different types of tissues with different staining techniques. Quantity, variability, and size of WSIs require efficient and versatile computer vision methods. As well, labeled data seems to be a rarity in digital pathology; manual delineation of regions of interest is not part of the clinical workflow, a fact that hinders the usage of many supervised learning algorithms.

The most actively researched task in digital pathology image analysis is the computer-assisted diagnosis (CAD), where the machine is trying to imitate  the pathologist's task. The diagnostic process maps a WSI or multiple WSIs to one of the disease categories. This is essentially a supervised learning task. Since the errors made by a computer system reportedly differ from those made by a human pathologist \cite{wang2016deep}, classification accuracy could be improved using the CAD system. The CAD may also lead to the reduced variability in interpretations and prevent overlooking regions of interest by investigating all pixels within WSIs.

Another research tasks in the digital pathology is Content-Based Image Retrieval (CBIR). A CBIR system  retrieves similar images when a query image is provided. In digital pathology, CBIR systems are useful in many situations, particularly in diagnostics, education, and research \cite{caicedo2011content,sridhar2015content}. For instance, CBIR systems can be used for educational purposes by students and pathologist residents to retrieve relevant cases or histopathology images. CBIR systems can help pathologists to diagnose rare cases if large archives are available.
 
Probably the biggest problem in  pathology image analysis using machine learning is that only a limited number  of labeled data is available. In the field of digital pathology, there are some public datasets that contain hand-labelled histopathology images \cite{gelasca2008evaluation,janowczyk2016deep,kumar2017comparative,xu2017large,gamper2019pannuke,shaban2020context,javed2020cellular,faust2018visualizing}. They could be useful if the purpose of the analysis, staining procedure, and magnification levels and resolutions are similar. However, because each of these datasets focuses on a specific disease, cell type, or staining, they are not generic enough to cover most necessary CAD tasks. There are also several large-scale histopathology image databases that contain high-resolution WSIs. The Cancer Genome Atlas (TCGA) \cite{weinstein2013cancer} contains over 30,000 WSIs from various cancer types, and the Genotype-Tissue Expression dataset (GTEx) \cite{lonsdale2013genotype} contains over 20,000 WSIs from various tissues. These datasets may serve as potential training data for various tasks. Furthermore, both TCGA and GTEx also provide genomic profiles, which could be used to investigate the relationships between genotype and morphology. 

One of pathology dataset is the Kimia Path24 dataset. 
This dataset includes 24 WSIs from different tissue textures and different staining procedures \cite{babaie2017classification}.
This dataset is structured to mimic retrieval tasks in clinical practice. The patch-to-scan classification may appear trivial but has relevant clinical applications such as ``\textbf{floater detection}'' where we need to find the origin of a foreign tissue \cite{floater1, floater2, pantanowitz2020digital}. To generate the training and test datasets, Babaie et al. \cite{babaie2017classification} slide a window with no overlapping over each WSI to crop patches of size 1000$\times$1000 and ignored background patches by analyzing both homogeneity and gradient change of each patch. The cropping generated a dataset with 28,380 patches.  

Color is an important characteristic in pathological images \cite{tosta2019computational,swiderska2020impact}. However, color has been ignored in the Kimia Path24 dataset, and all patches are saved as grayscale images after being extracted from RGB WSIs. To tackle this problem, we provide a new dataset with RGB patches from  the original WSIs in the Kimia Path24 dataset. We call this dataset “\emph{Kimia Path24C}”. To provide preliminary results and set a benchmark for the proposed dataset, we used three pre-trained deep networks for feature extraction. Subsequently, we use these feature vectors for image retrieval. We show that adding color to the dataset combined with using deep features provide high retrieval accuracy.  
 This data set is publicly available on the website of Kimia Lab\footnote[1]{\protect\url{https://kimia.uwaterloo.ca/}}.
 
\section{Related Works}
\subsection{Convolutional Neural Networks}
A convolutional neural network (CNN) is a type of deep, feed-forward artificial neural network which receives raw input images and extracts complex features from them to output class assignments \cite{simard2003best}. The CNNs use the convolution of spatial information in the images to share weights across units.  

Most of the image recognition techniques have been replaced by deep learning algorithms, after the outstanding results of deep learning algorithms in the analysis of large-scale databases \cite{krizhevsky2012imagenet}.
More recently, researchers have been working on applying deep learning to pathology image analysis. For example, Spanhol et al. \cite{spanhol2016breast} present a set of experiments using a deep learning approach to avoid hand-crafted features. They have shown that an existing CNN architecture designed for classifying color images of objects can be used for the classification of histopathology images. Xu et al. \cite{xu2016deep} proposed a deep convolutional neural network based on feature learning to automatically segment or classify epithelium and stroma regions from digitized tumor tissue microarrays. Automated breast cancer multi-classification using histopathological images is of great clinical significance.  Han et al. \cite{han2017breast} proposed a breast cancer multi-classification method using a deep learning model that achieves remarkable performance (average 93.2\% accuracy) on a large dataset. Mormont et al. 
\cite{mormont2018comparison} study transfer learning as a way of overcoming object recognition challenges encountered in the field of digital pathology. Their experiments on eight classification datasets show that densely connected and residual networks consistently yield best performances across all topologies. 

Khosravi et al. \cite{khosravi2018deep} introduced a classification pipeline including a basic CNN architecture, Google's Inceptions with three training strategies, and an ensemble of Inception and ResNet to effectively classify different histopathology images for different types of cancer. They demonstrated the accuracy of deep learning approaches for identifying cancer subtypes, and the robustness of Google’s Inceptions even in presence of extensive tumor heterogeneity. On average, their pipeline achieves accuracy values of 100\%, 92\%, 95\%, and 69\% for discrimination of various cancer tissues, subtypes, biomarkers, and scores, respectively. For more information about deep neural network approaches and their application in medical image analysis see Litjens et al. \cite{litjens2017survey}.

\subsection{Segmentation}
Segmentation is a common task in both natural and medical image analysis. K-means clustering and Gaussian Mixture Model (GMM) appear to be popular choices for the unsupervised segmentation tasks. The K-means clustering is an iterative distance-based clustering algorithm that assigns $n$ observations to exactly one of the $k$ clusters defined by centroids, where $k$ is chosen before the algorithm starts \cite{hartigan1979algorithm}. The Gaussian mixture model is a well-known method used in most applications such as data mining, pattern recognition, machine learning, and statistical analysis. GMM assumes the data to consist of several Gaussian distributions, which can be discovered through the parameter learning process. The major advantage of the GMM approach is that its mathematical equation is easy to evaluate. As well, it requires a small number of parameters for learning. For more information on GMM see McLachlan and Peel \cite{mclachlan2004finite}.

Image segmentation is one of the main components in a fully automated cell image analysis. Segmentation mainly focuses on the separation of the cells from the background as well as separation of the nucleus from the cytoplasm within the cell regions for extracting cellular features from images. For example,  Bamford and Lovell \cite{bamford1998unsupervised} segmented the nuclei in a pap smear image using an active contour model that was estimated by using dynamic programming to find the boundary with the minimum cost within a bounded space around the darkest point in the image. Yang-Mao et al. \cite{yang2008edge} applies automatic thresholding to the image gradient to identify the edge pixels corresponding to nuclei and cytoplasm boundaries in cervical
cell images. Tsai et al. \cite{tsai2008nucleus} replace the thresholding step with K-means clustering into two partitions. Ragothaman et al. \cite{ragothaman2016unsupervised} used the GMM to segment cell regions to identify cellular features such as the nucleus and cytoplasm. Segmentation is a crucial task for WSI representation. In this paper, we utilize both K-means and GMM to localize the tissue region and extract patches from the informative part of the WSI and avoid background pixels.  

\section{A new dataset: Kimia Path24C}
In this paper, we assemble, test and propose a new version of the Kimia Path24 dataset introduced by Babaie et al. \cite{babaie2017classification}. It contains 24 WSIs with several staining techniques namely immunohistochemical (IHC), Hematoxylin and Eosin (H\&E) and Masson's trichrome staining, selected from 350 WSIs depicting diverse body parts so that the images clearly represent different texture pattern \cite{babaie2017classification}. 
Babaie et al. \cite{babaie2017classification} have visually selected 1,325 patches of size 1000$\times$1000 at 20$\times$ magnification as test dataset that represent all dominant tissue textures in each WSI. Indeed,  test dataset contains multiple texture patterns.

Considering Babaie et al. \cite{babaie2017classification} saved the extracted patches as grayscale images, the staining information that is very important in the analysis of histopathological images was lost. 
We create a new version of the dataset that not only contains the staining information but also contains less irrelevant patches (i.e., patches with background pixels and debris). Besides, we would like to fix the training patches such that benchmarking becomes more consistent.

\subsection{Patch extraction}
In the new version of the Kimia Path24 dataset, the locations and number of patches of the test dataset have not changed and only the extracted patches have been saved as RGB images instead of grayscale. 
For patch extraction and creating an informative training dataset, we first identify tissue within each WSI and exclude background pixels (i.e., largely white/bright space). Background detection is needed to reduce the computation time and find informative regions that contain tissue patterns suitable for analysis. To achieve this, we use two popular segmentation algorithms, namely the K-means and GMM, to automatically detect the background pixels.
In particular, all thumbnails were first extracted in $1\times$ magnification and artifacts such as air bubbles under the cover slip or dust were manually eliminated. Then each thumbnail is segmented into five clusters, namely the background, fat, cell nuclei, connective tissues, and blood. After segmentation, a window is moved over each label matrix to crop patches of size 50$\times$50 pixels in $1 \times$ magnification. The label of background pixels assigned by the segmentation algorithm is found using the patch in the top right corner of the label matrix. If background share of a patch is less than a predefined value, we consider that patch as a suitable patch (containing enough tissue sample) otherwise we ignore it. Finally, using the coordinates of selected patches, we extract the corresponding patches of size 1000$\times$1000 from 20$\times$ magnification level. Note that to construct a training dataset, we first remove (whiten) locations of test patches in each WSI so that these patches cannot be mistakenly used for training.

\subsection{Deep networks}
Extracting suitable features is a critical step for successfully implementing an image search or classification algorithm. The literature in recent years suggests that the pre-trained deep networks can be used as fine feature extractors \cite{kieffer2017convolutional}. A significant benefit of the pre-trained deep networks is that we do not need to train a deep network again and extracted features can be directly applied to the existing image analysis pipelines \cite{litjens2017survey}. 

In this study, we used three pre-trained deep networks, namely \textit{VGG16} \cite{simonyan2014very}, \textit{InceptionV3} \cite{szegedy2016rethinking},  and \textit{DenseNet-121} \cite{huang2017densely}, as feature extractors without any fine-tuning. These networks have been trained on more than a million images from the \emph{ImageNet} database \cite{russakovsky2015imagenet} and can classify images into 1000 object categories, such as pencil, keyboard, and many objects and animals. 
Simonyan et al. \cite{simonyan2014very} proposed \textit{VGG16} architecture with $16$ layers being deeper than the other existing architectures at that time. They also utilized $3 \times 3$ kernels with a stride of $1$ pixel in order to decrease the number of parameters. \textit{InceptionV3} is a 42-layer network introduced by Szegedy et al. \cite{szegedy2016rethinking}.  They believed that the use of multiple convolutional layers with a small filter size instead of a single convolutional layer with a large filter size can reduce the number of parameters and the computational cost, and can still achieve a similar level of model expressiveness.
By the introduction of the \textit{DenseNet} by Huang et al. \cite{huang2017densely}, it was shown that deep networks can achieve the highest accuracy values while they stay efficient if they have shorter connections between layers either close to the input or the output of the network. Unlike the original neural network configurations where an $L$-layer network has $L$ connections, in a \textit{DenseNet} the feature maps of all preceding layers are fed into each layer. This solution not only diminishes the vanishing gradient problem but also enhances feature propagation and feature reuse, while it decreases the number of parameters.

We run each experiment using the architecture for the \textit{VGG16}, \textit{Inceptionv3} and \textit{DenseNet} networks as provided in the Keras Python package \cite{chollet2015keras}. The final average pooling layer right before the last fully connected layer of these networks prior to classification was extracted as a feature vector. The length of the feature vector from \textit{VGG16}, \textit{Inceptionv3} and \textit{DenseNet} is  512, 2048 and 1024, respectively. 


\section{Results and discussion}
\subsection{Experimental results}
To determine the optimal size of the training dataset necessary to achieve high accuracy, we consider seven different candidate values for the background. In addition, for each of the 24 WSIs, patches are extracted with different overlap ratios, without any overlap, and with 20\% overlap. Overall, we extracted 28 different training datasets using these strategies. The number of extracted patches for the training dataset via different strategies used for performance evaluation is reported in Table \ref{tab1}.


\begin{table}[htb!]
\begin{center}
\caption{The number of extracted patches via different strategies.}
\label{tab1}
\resizebox{\columnwidth}{!}{
\begin{tabular}{lcccccccccc}
\hline
& \multicolumn{2}{c}{\textbf{K-means}} & \multicolumn{3}{c}{\textbf{GMM}}   \\
\cline{2-3}  \cline{5-6} 
\backslashbox[28mm]{Back. \%}{Overlap \%}  & 0 & 20 && 0  & 20 \\
\hline
10 & 11429 & 18795 && 13425 & 23144  \\   
20 & 13293 & 21900 && 15636 & 26155  \\   
30 & 14804 & 24412 && 17750 & 29125  \\   
40 & 16313 & 26987 && 19479 & 31924  \\   
50 & 17888 & 29558 && 20752 & 34209  \\   
60 & 19437 & 32136 && 21811 & 35944  \\   
70 & 20927 & 34549 && 22591 & 37444  \\   
\hline
\end{tabular}
}
\end{center}
\end{table}

\begin{figure*}[htb]
\begin{center}
\includegraphics[height=3cm, width=3.5cm]{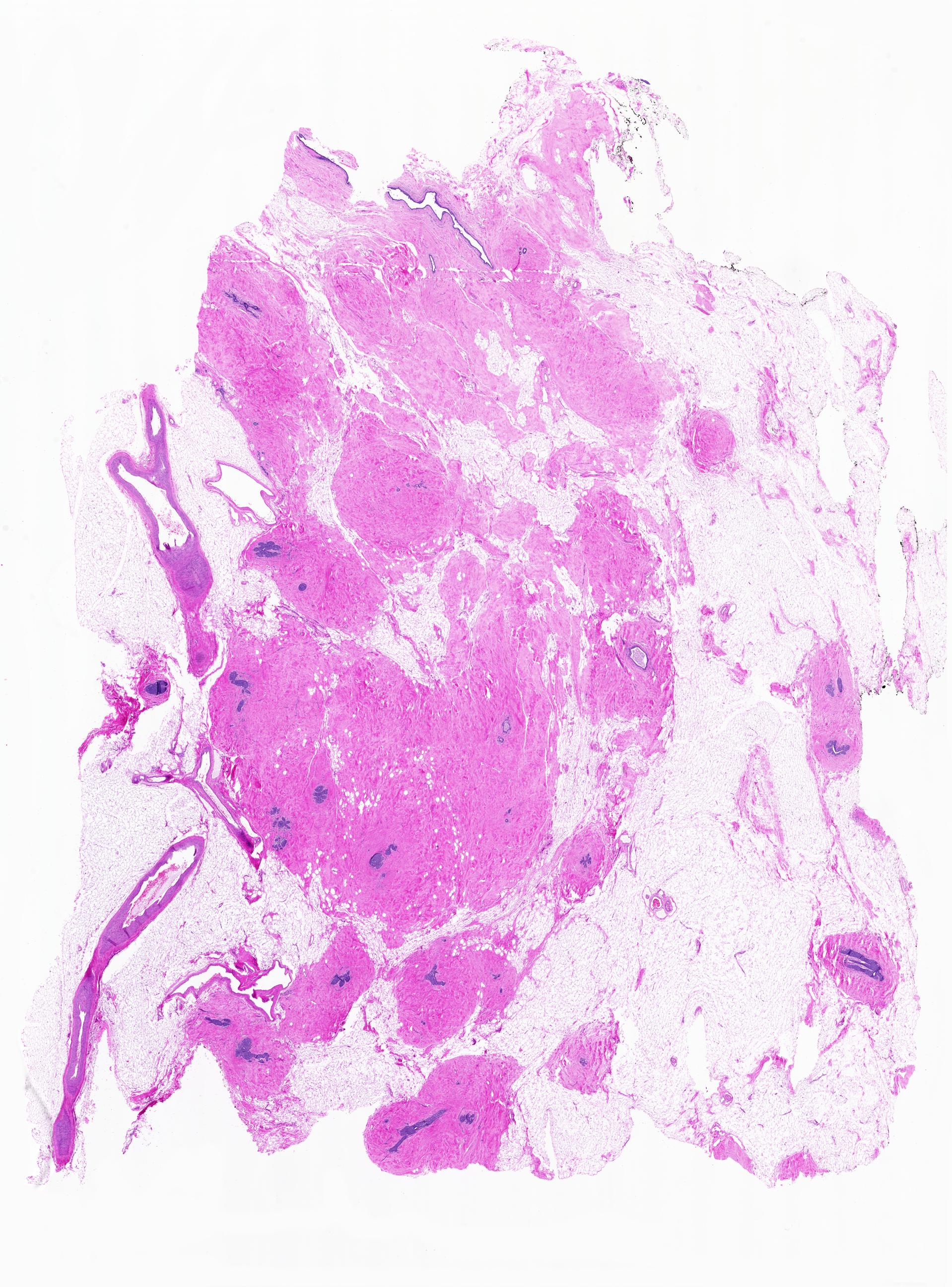}
\includegraphics[height=3cm, width=3.5cm]{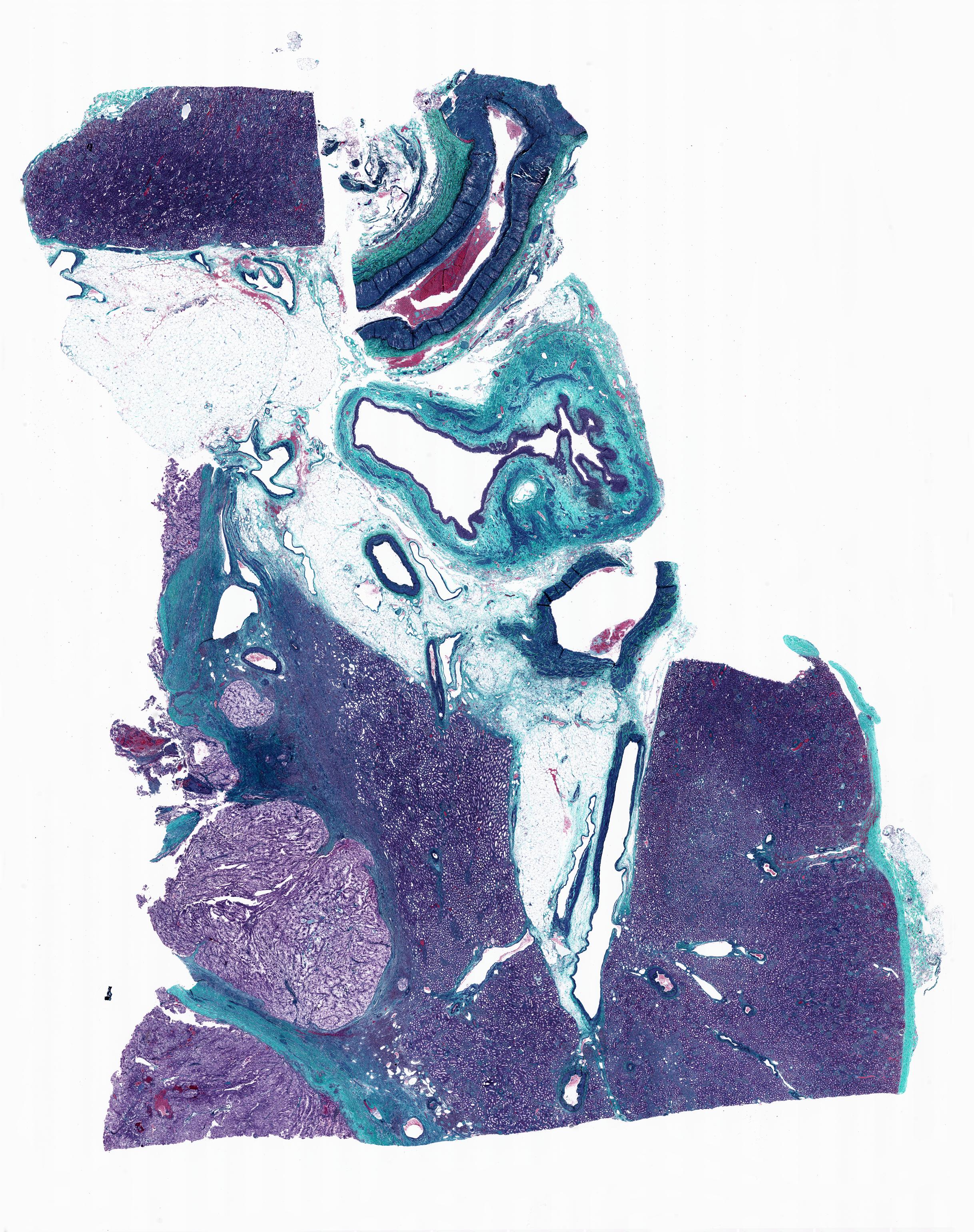}
\includegraphics[height=3cm, width=3.5cm]{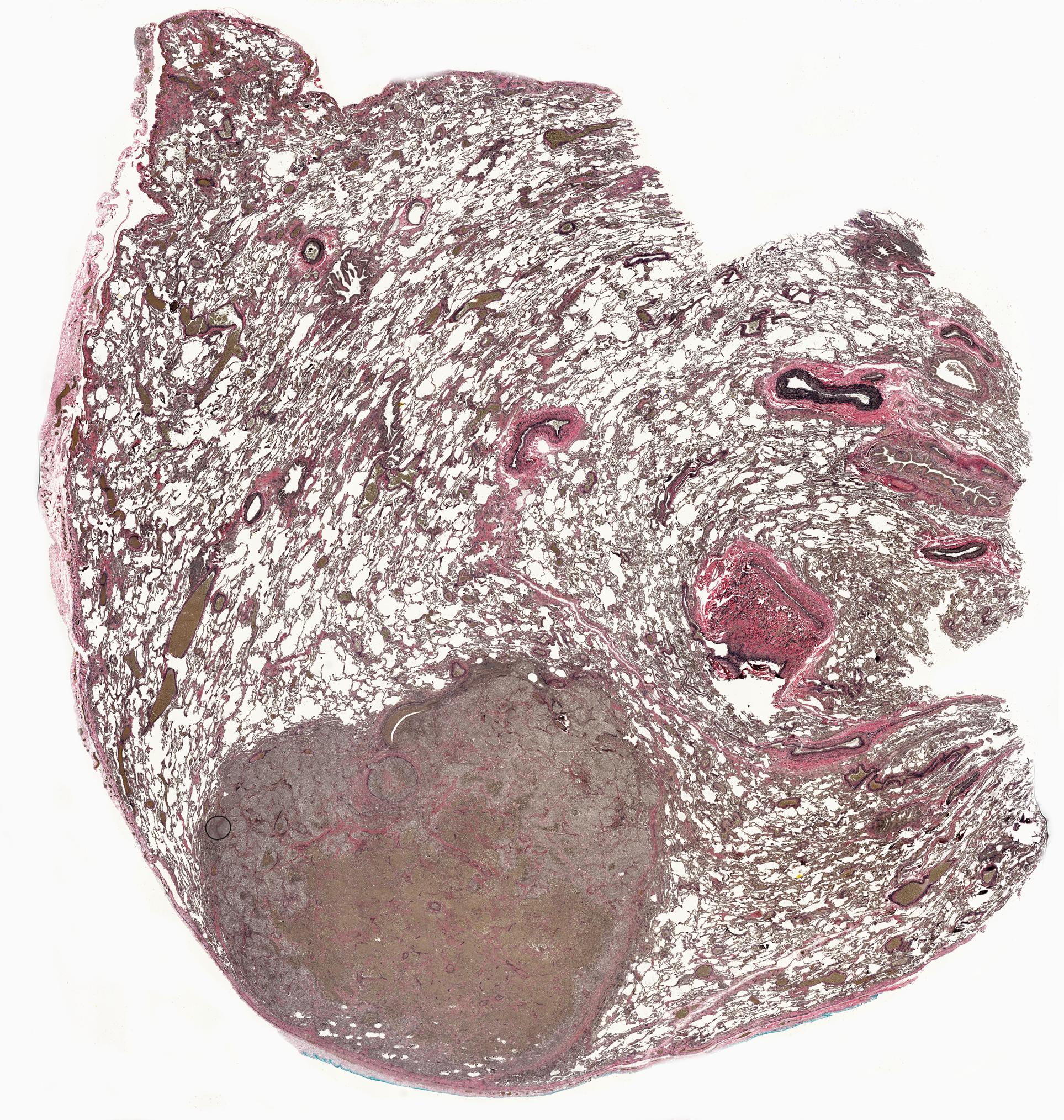}
\includegraphics[height=3cm, width=3.5cm]{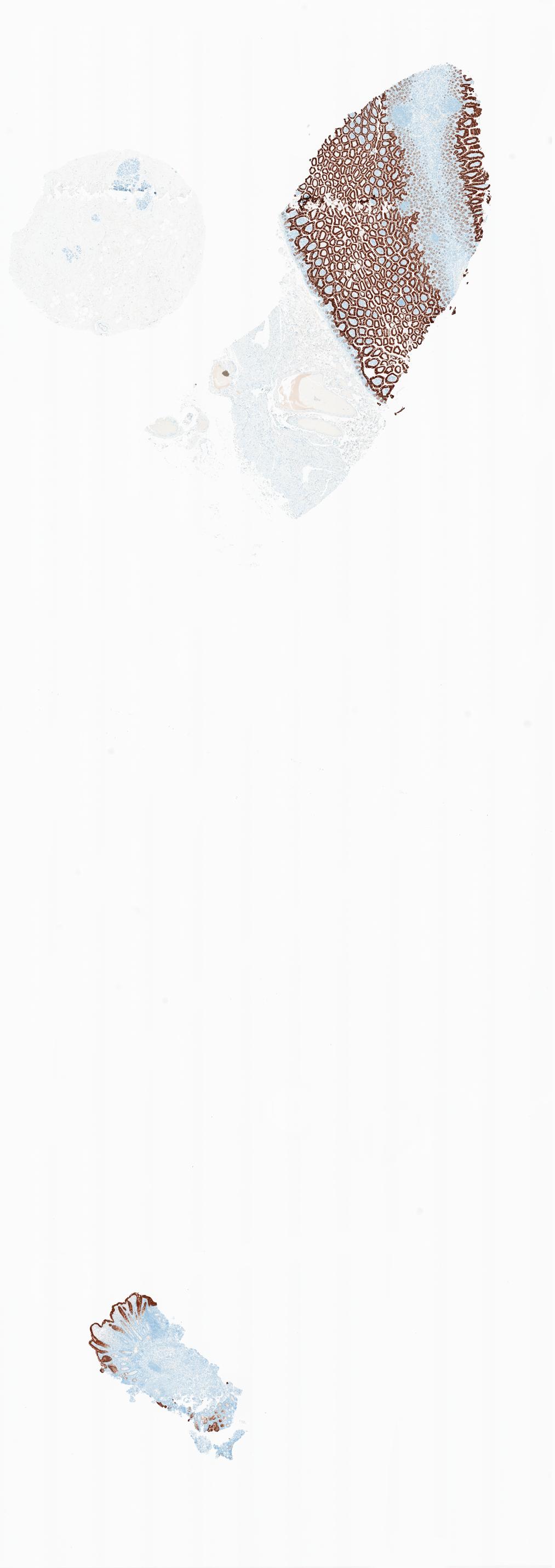}
\includegraphics[height=3cm, width=3.5cm]{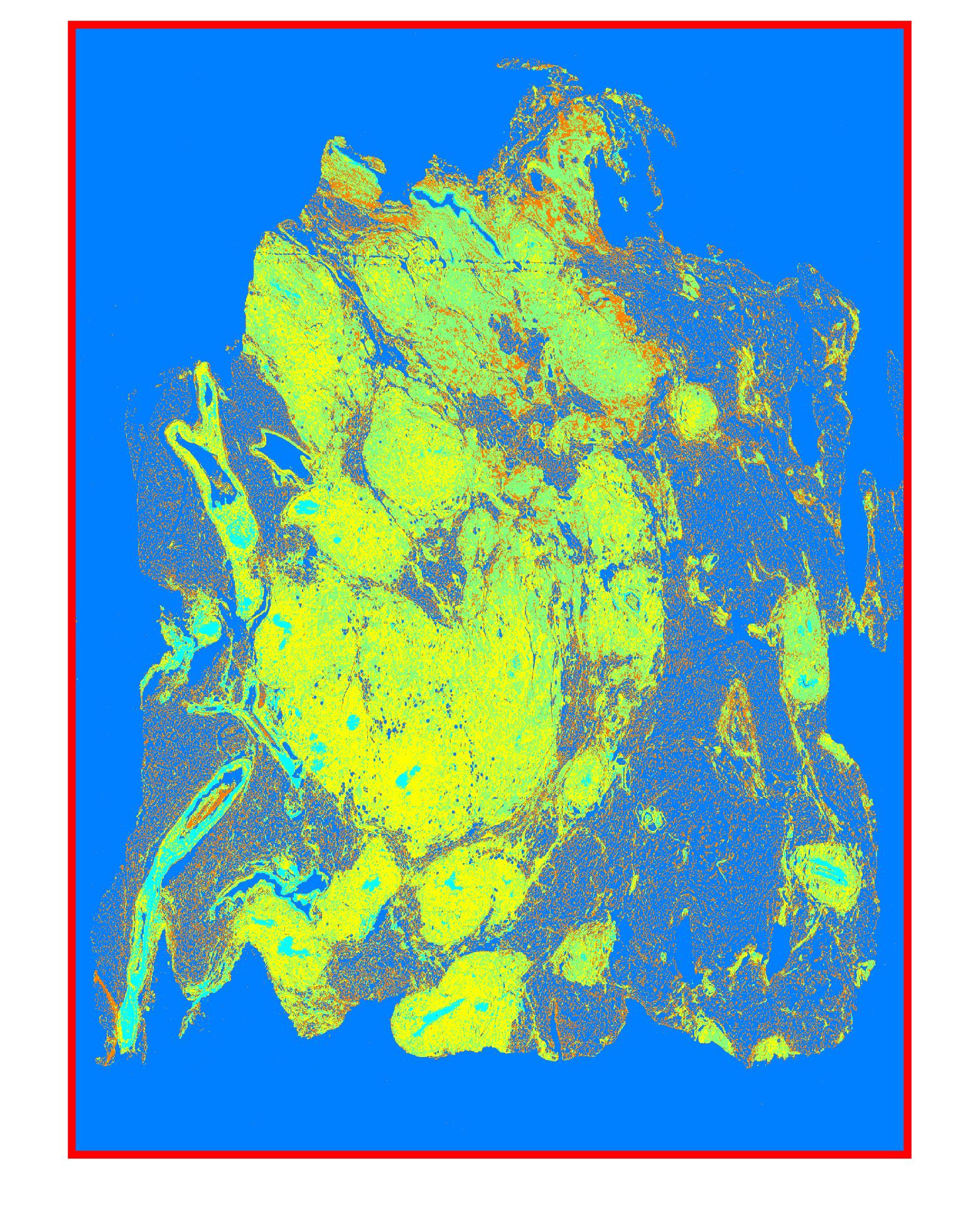}
\includegraphics[height=3cm, width=3.5cm]{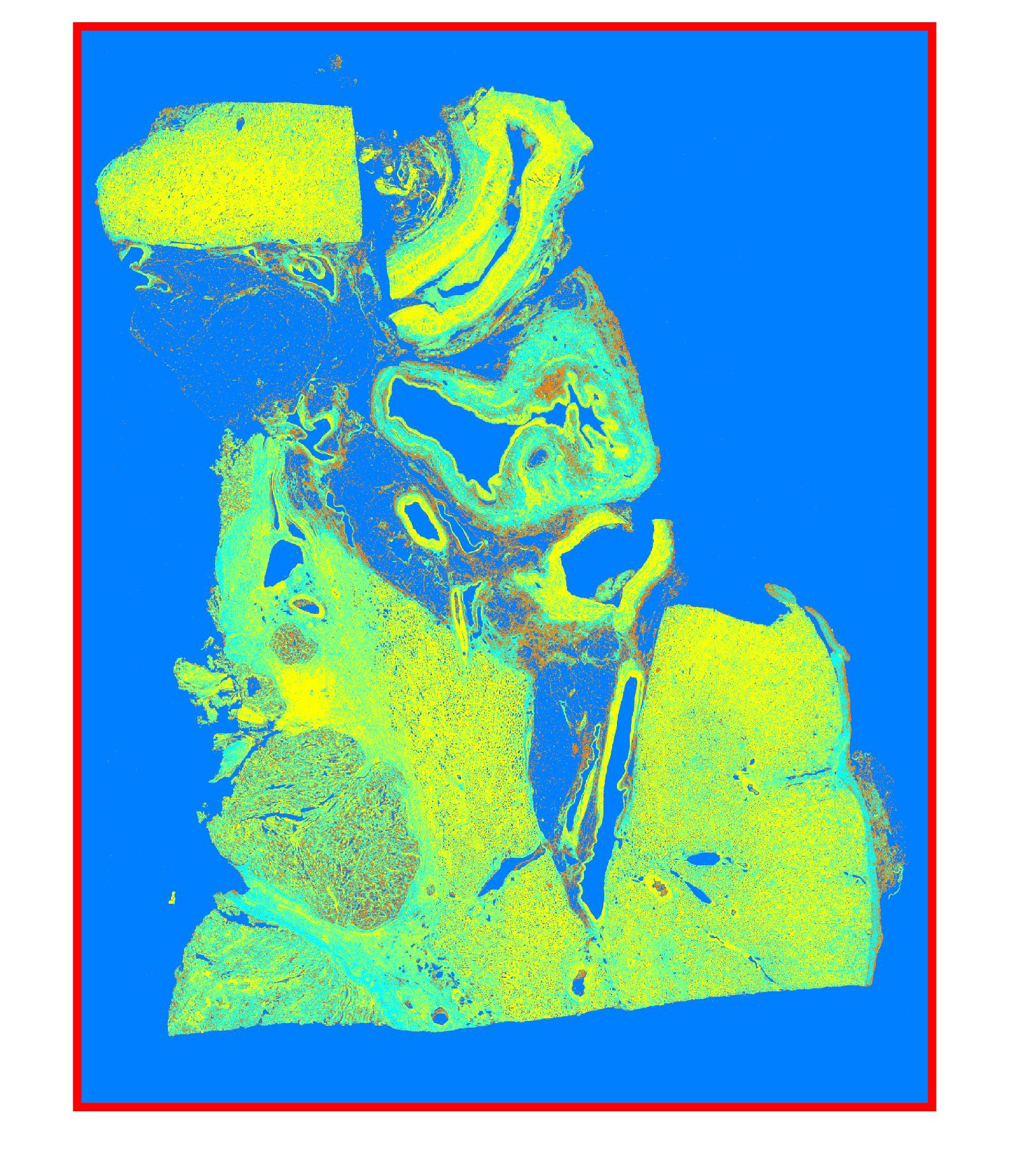}
\includegraphics[height=3cm, width=3.5cm]{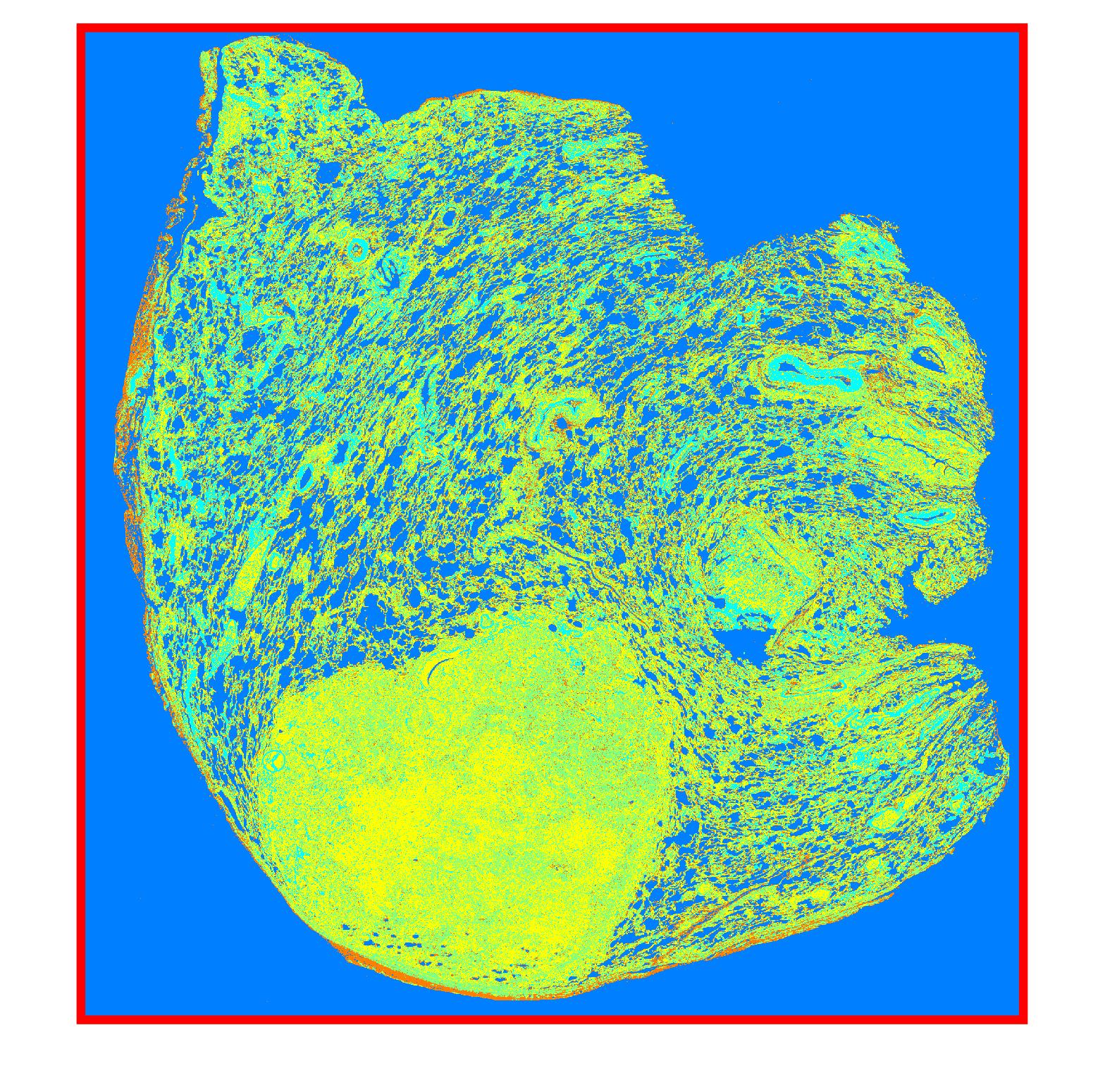}
\includegraphics[height=3cm, width=3.5cm]{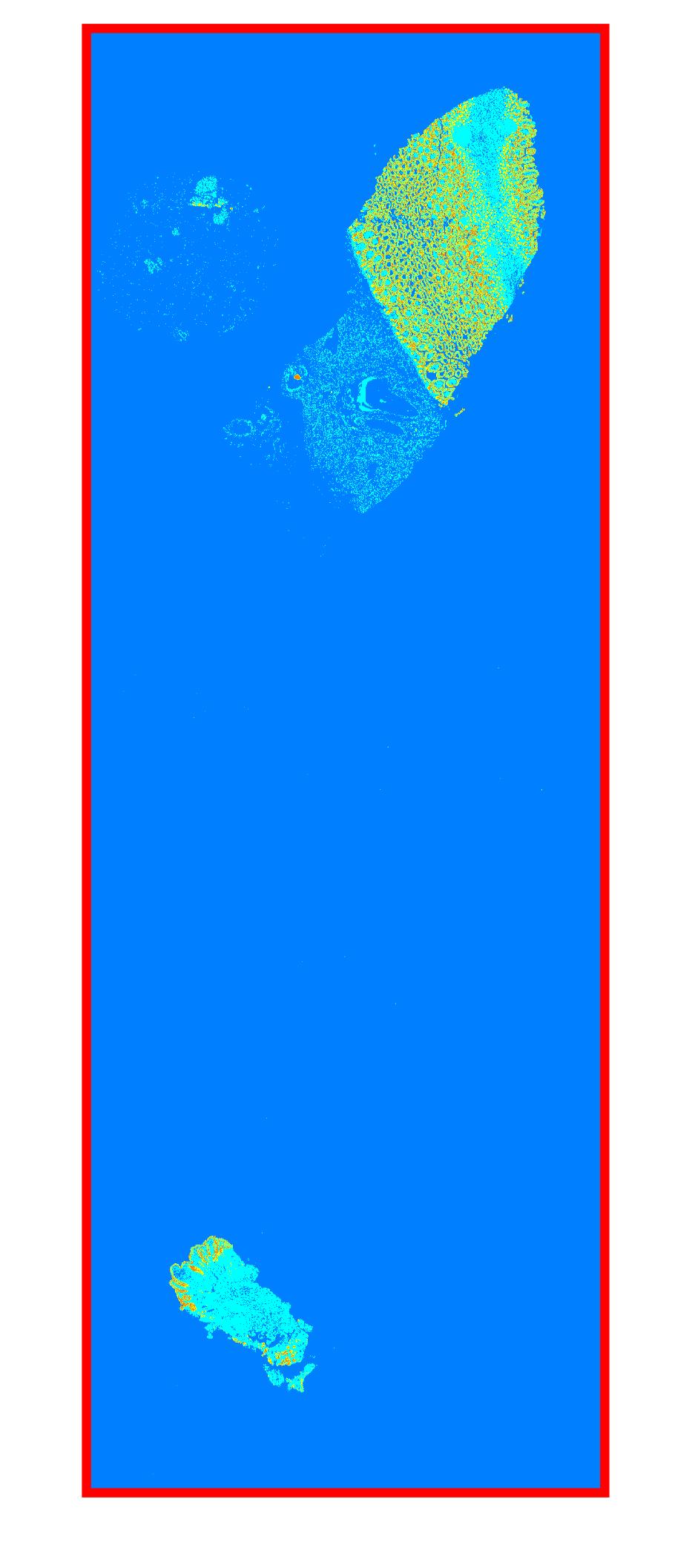}
\includegraphics[height=3cm, width=3.5cm]{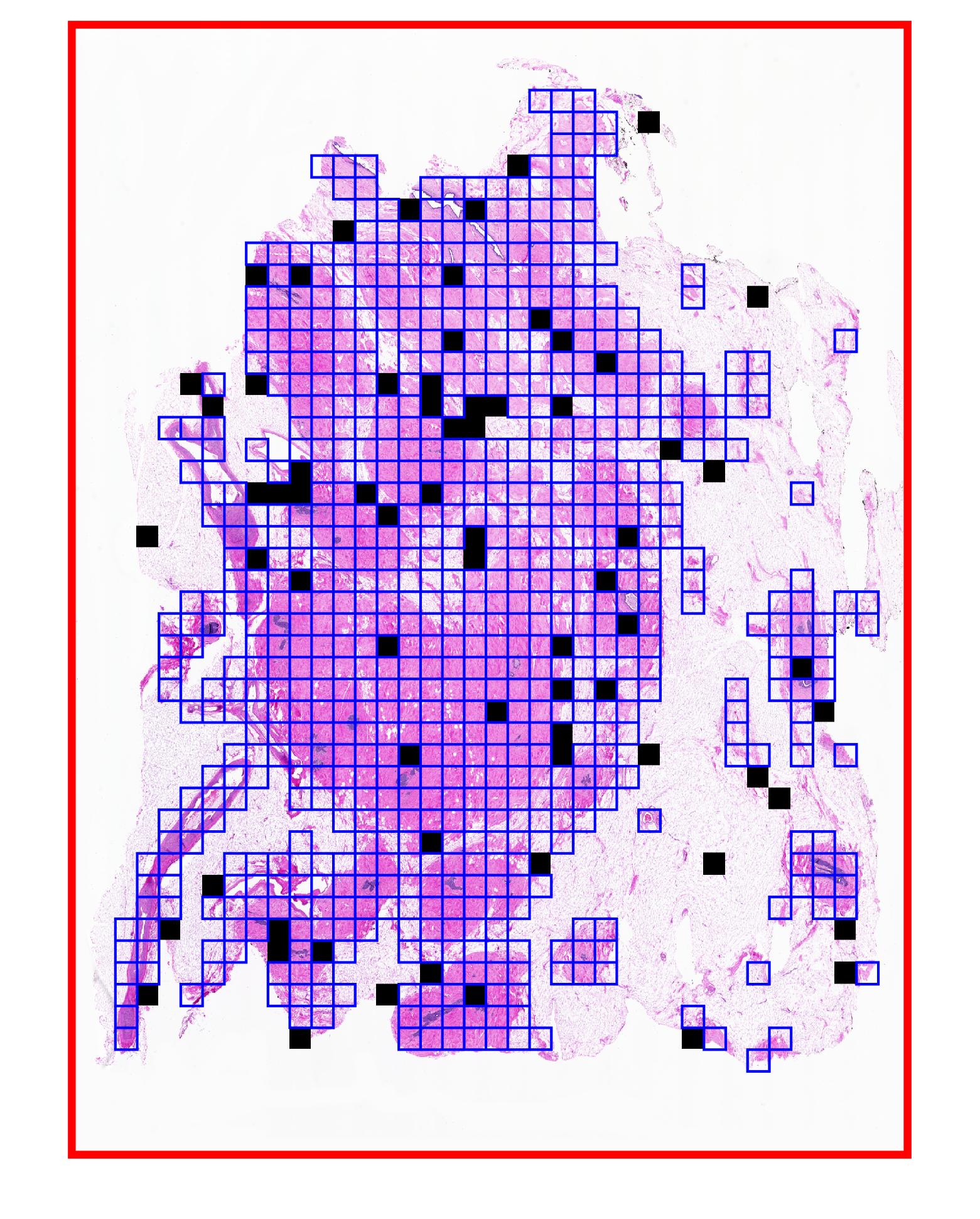}
\includegraphics[height=3cm, width=3.5cm]{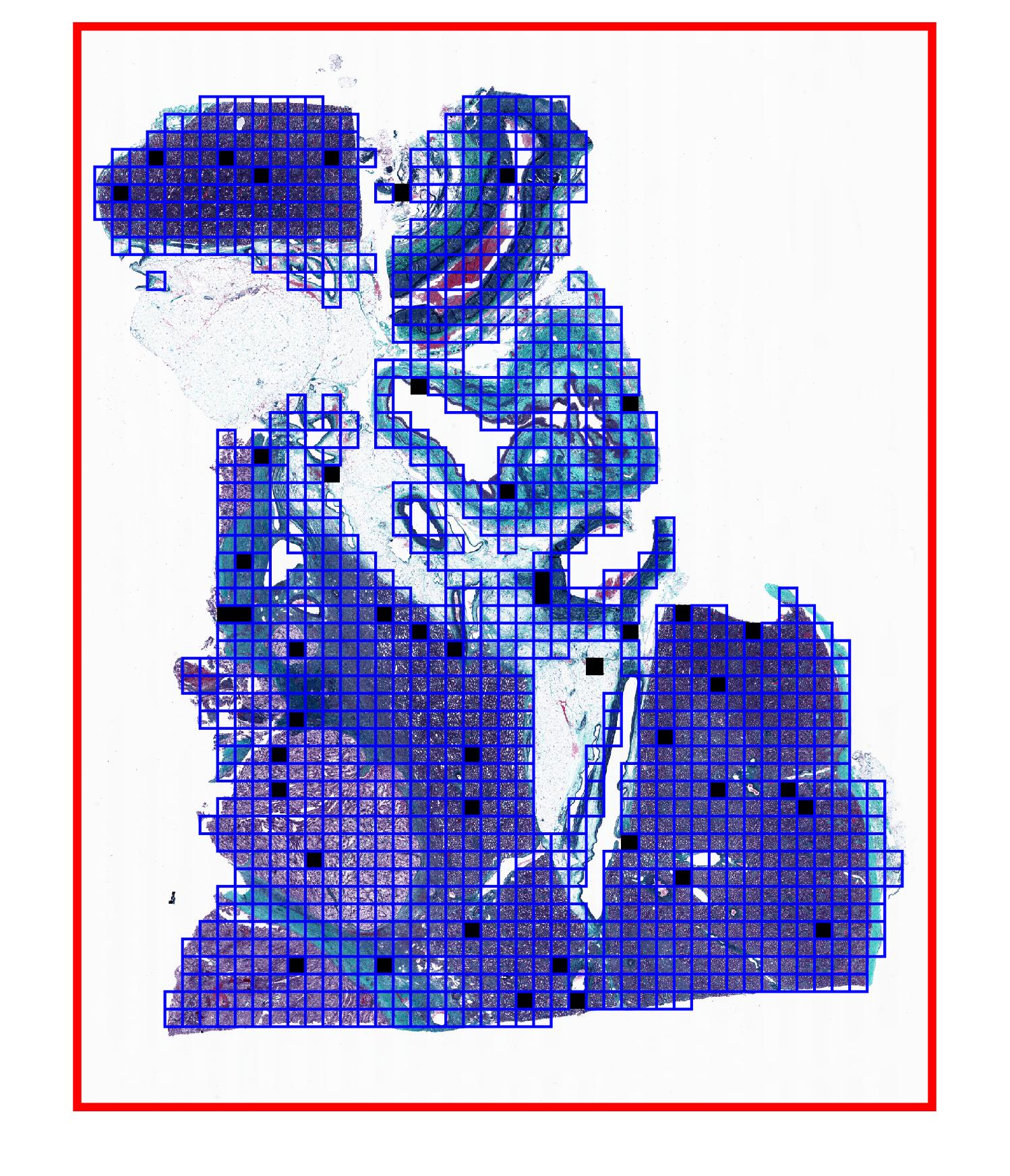}
\includegraphics[height=3cm, width=3.5cm]{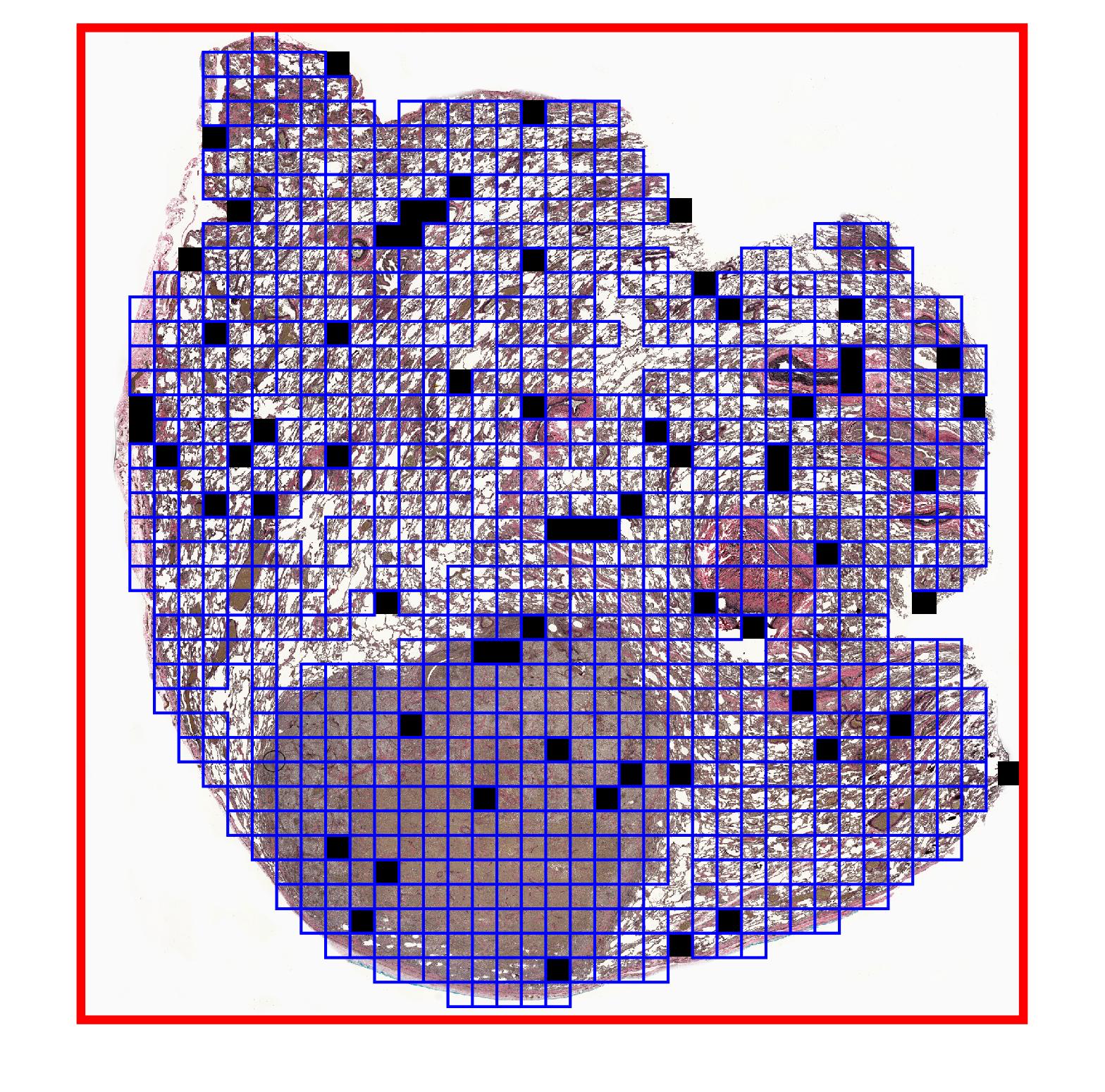}
\includegraphics[height=3cm, width=3.5cm]{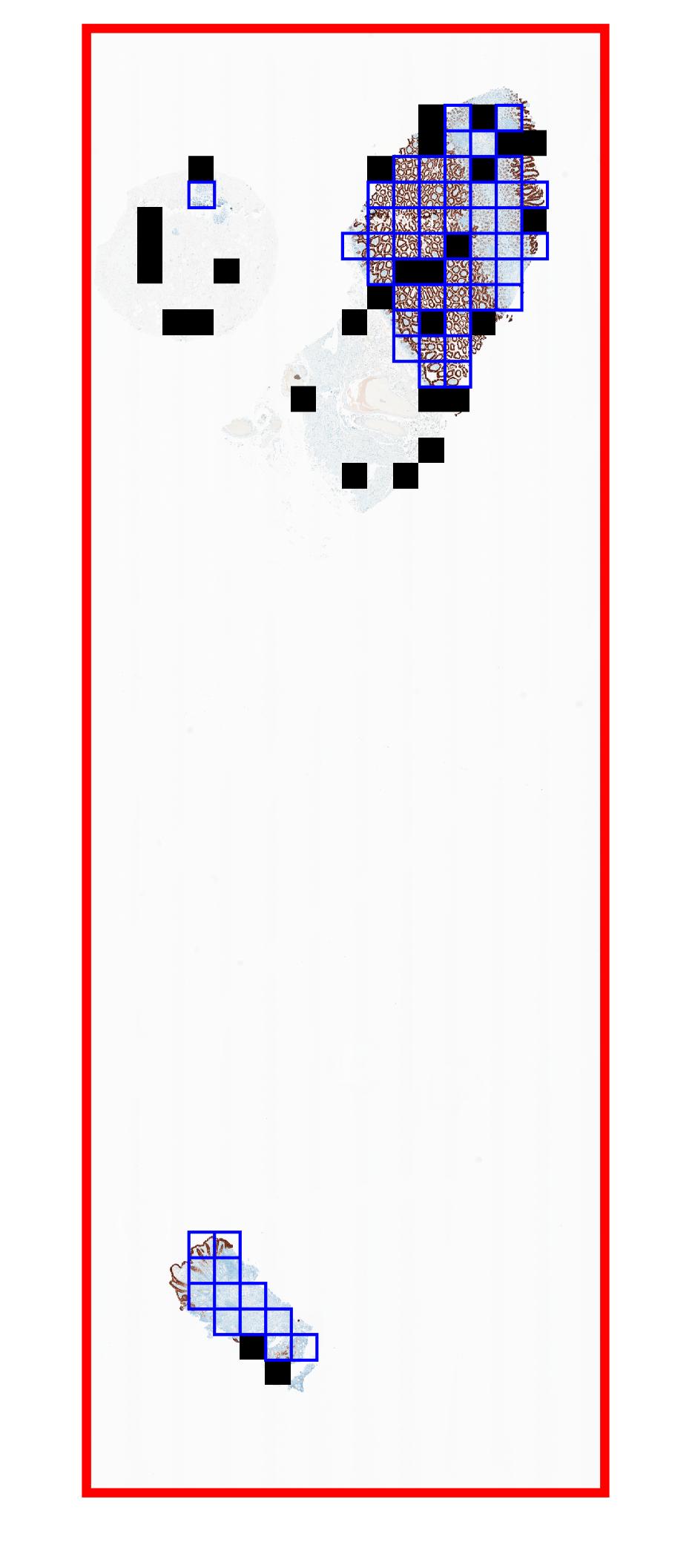}
\includegraphics[height=3cm, width=3.5cm]{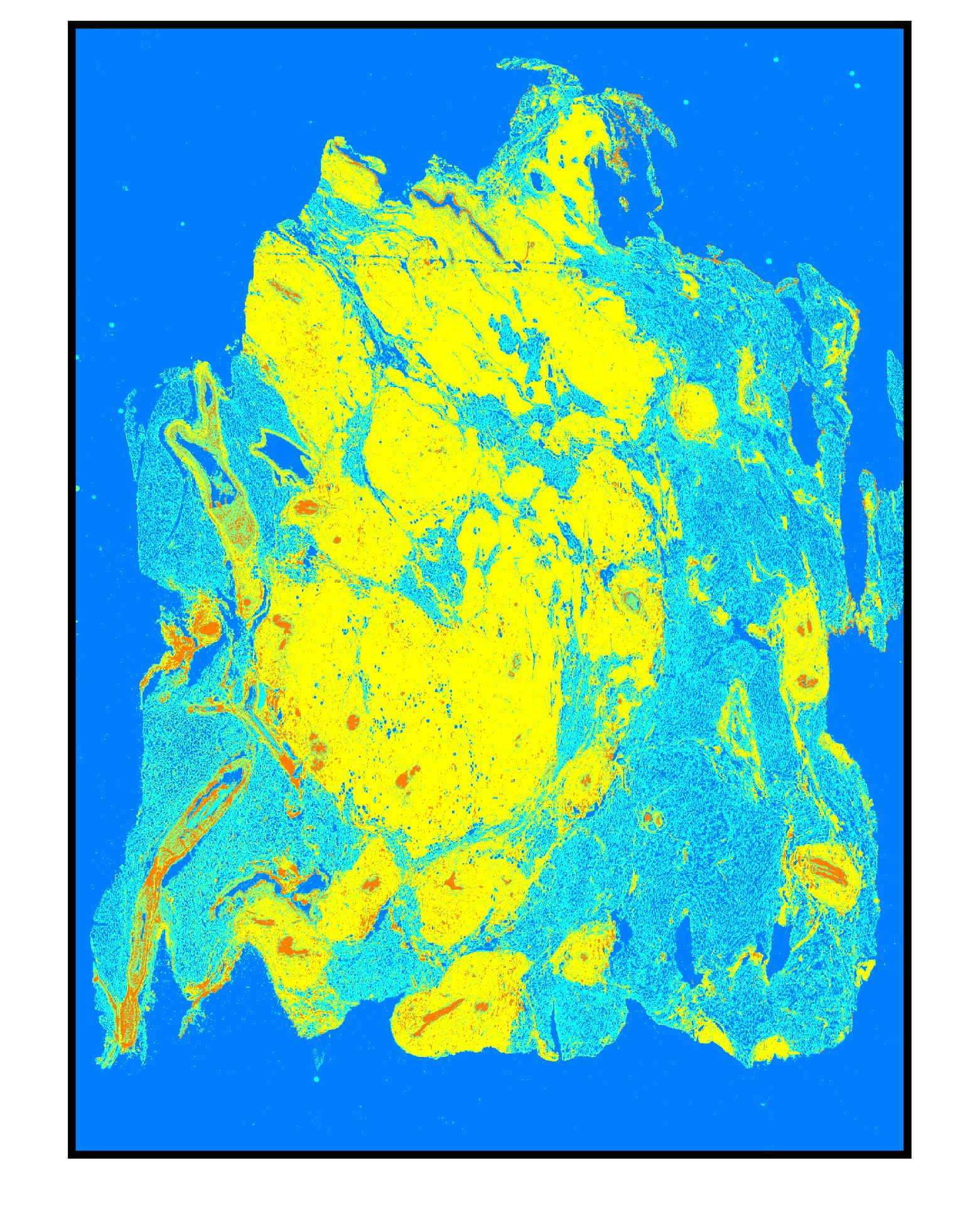}
\includegraphics[height=3cm, width=3.5cm]{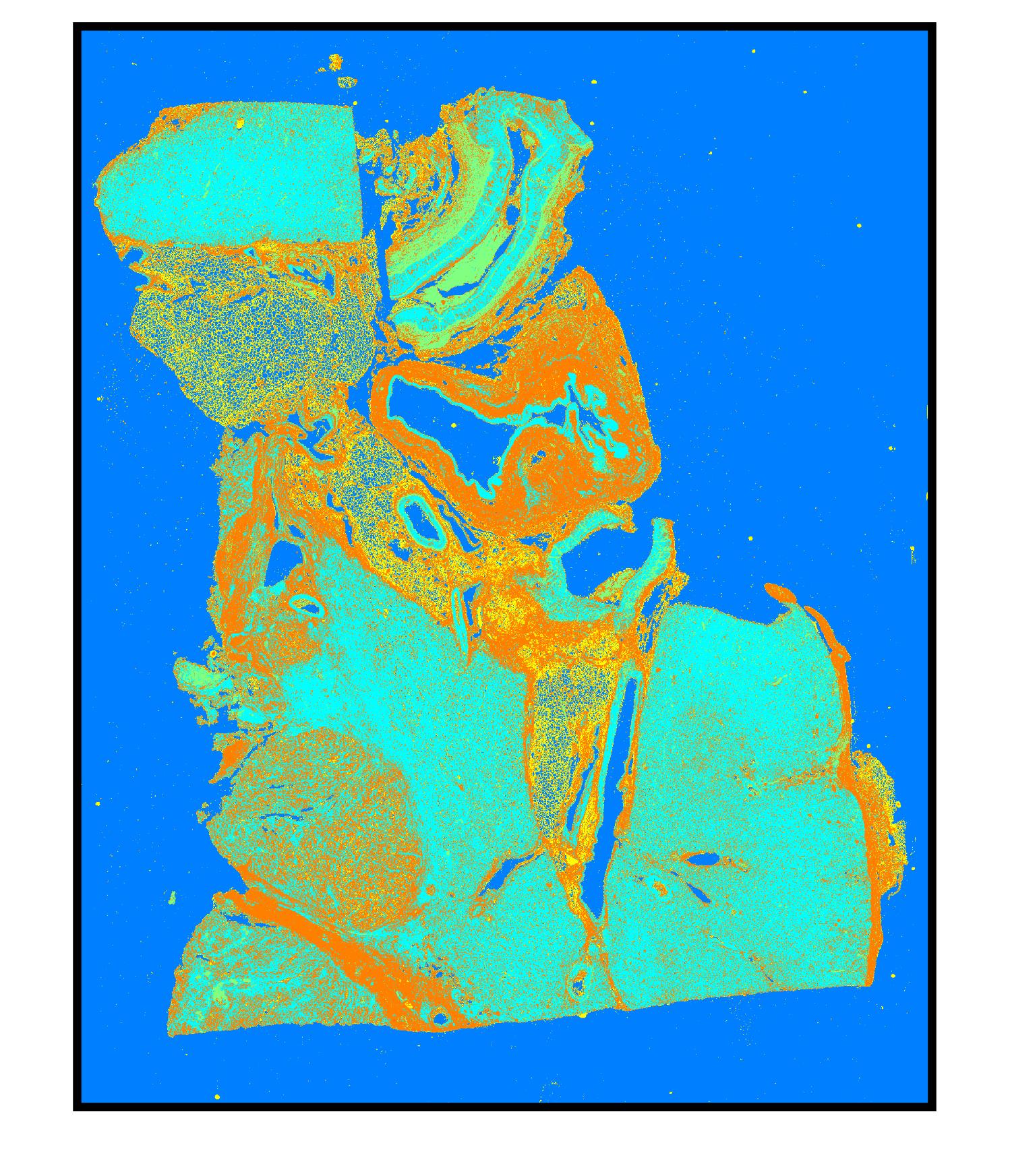}
\includegraphics[height=3cm, width=3.5cm]{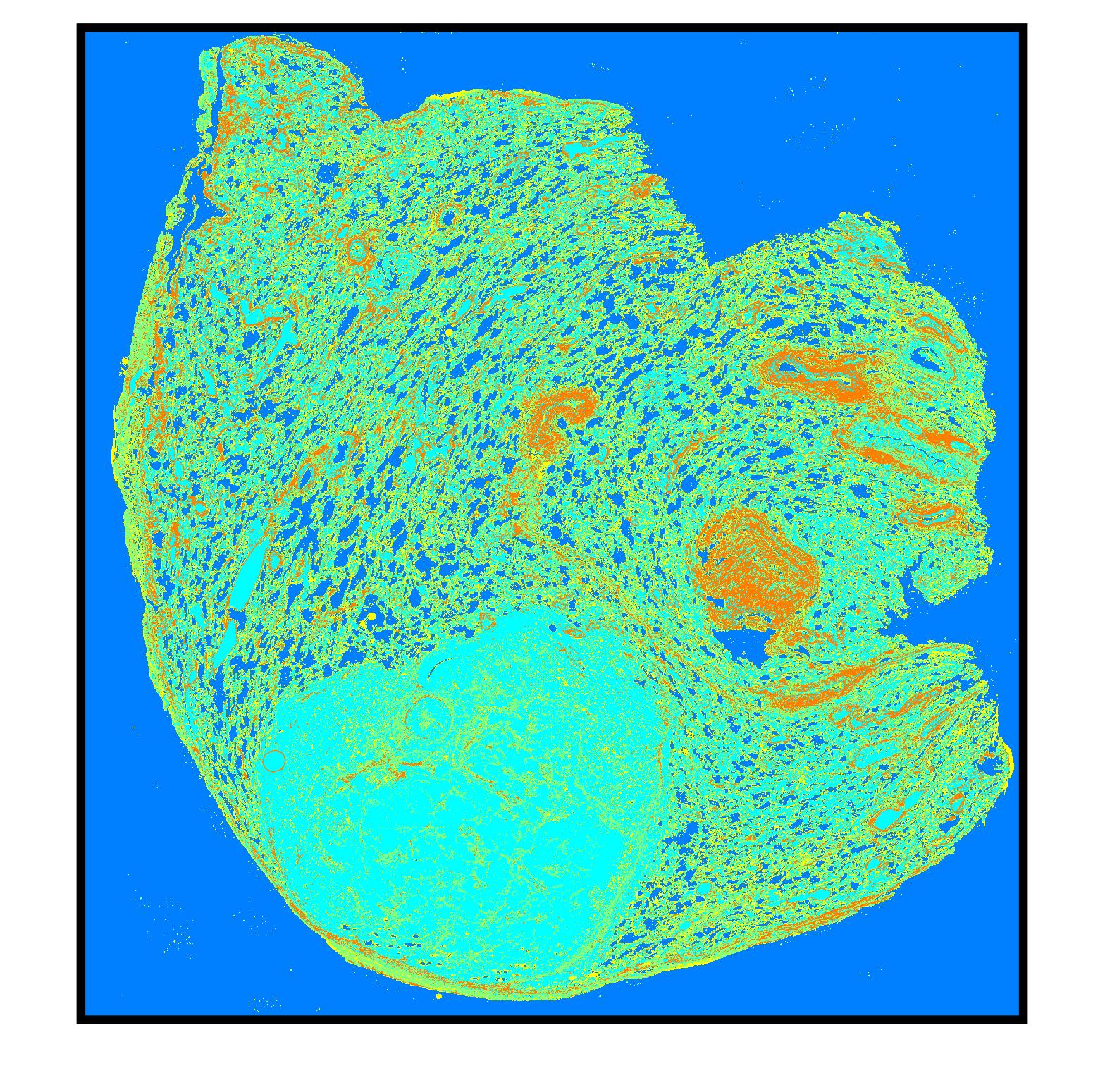}
\includegraphics[height=3cm, width=3.5cm]{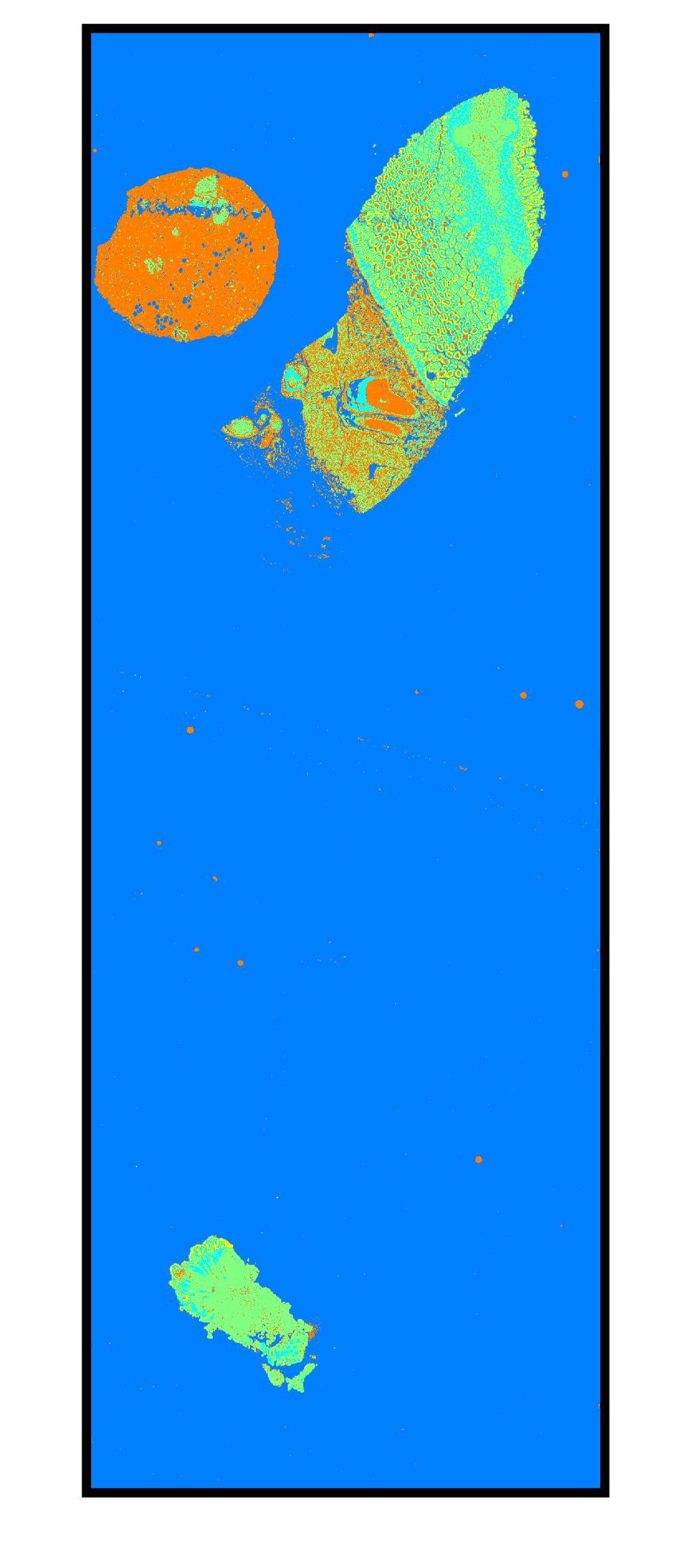}
\includegraphics[height=3cm, width=3.5cm]{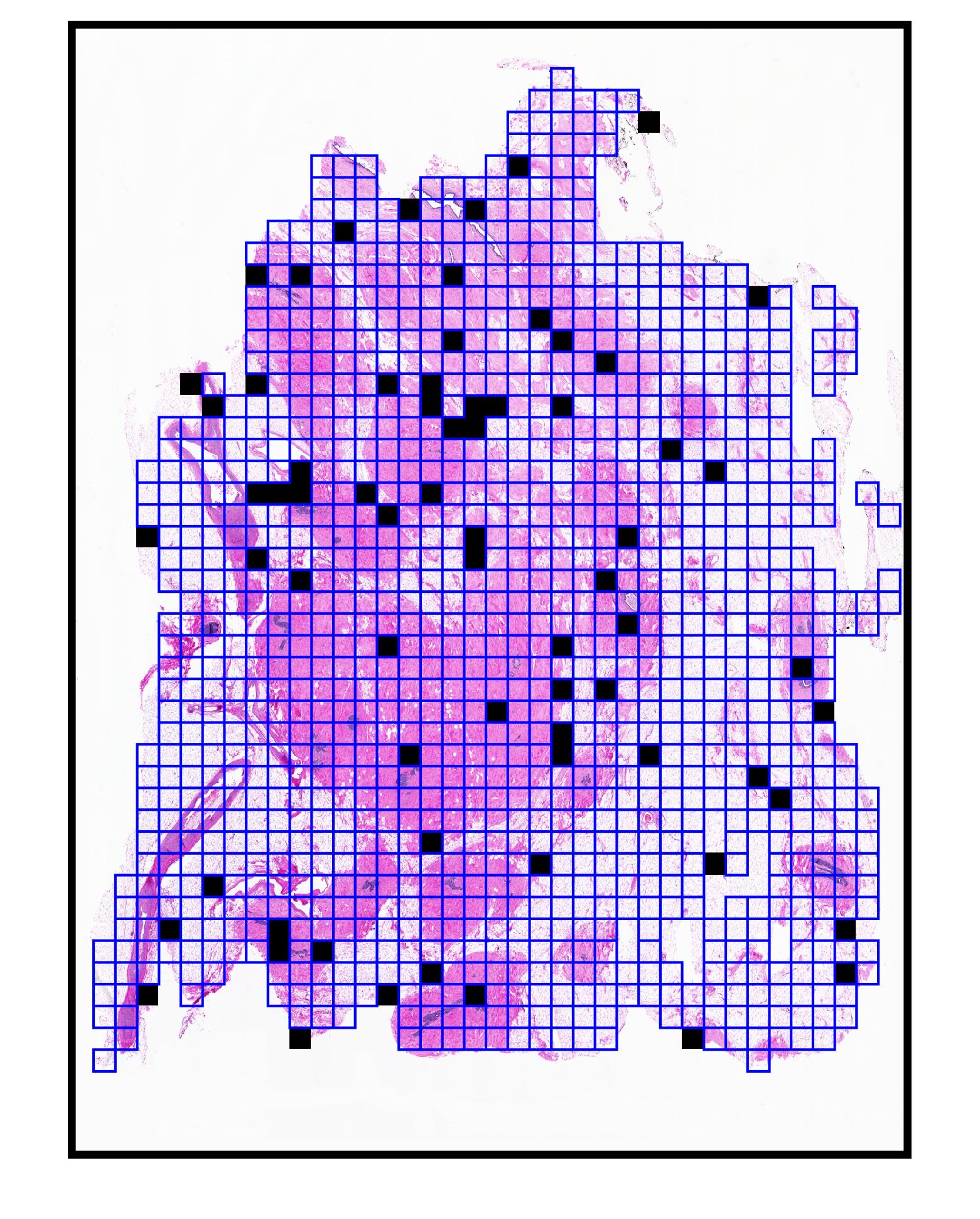}
\includegraphics[height=3cm, width=3.5cm]{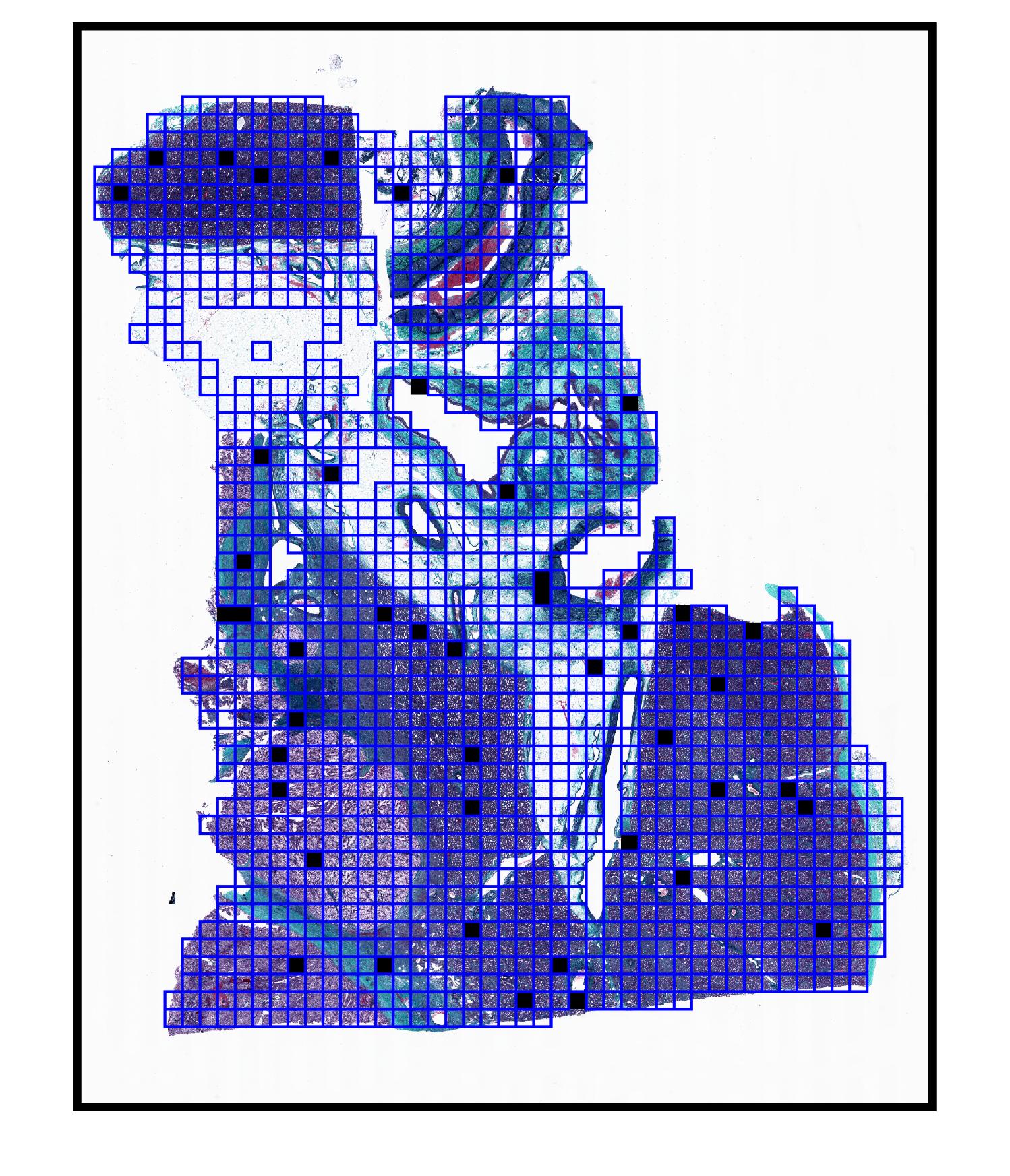}
\includegraphics[height=3cm, width=3.5cm]{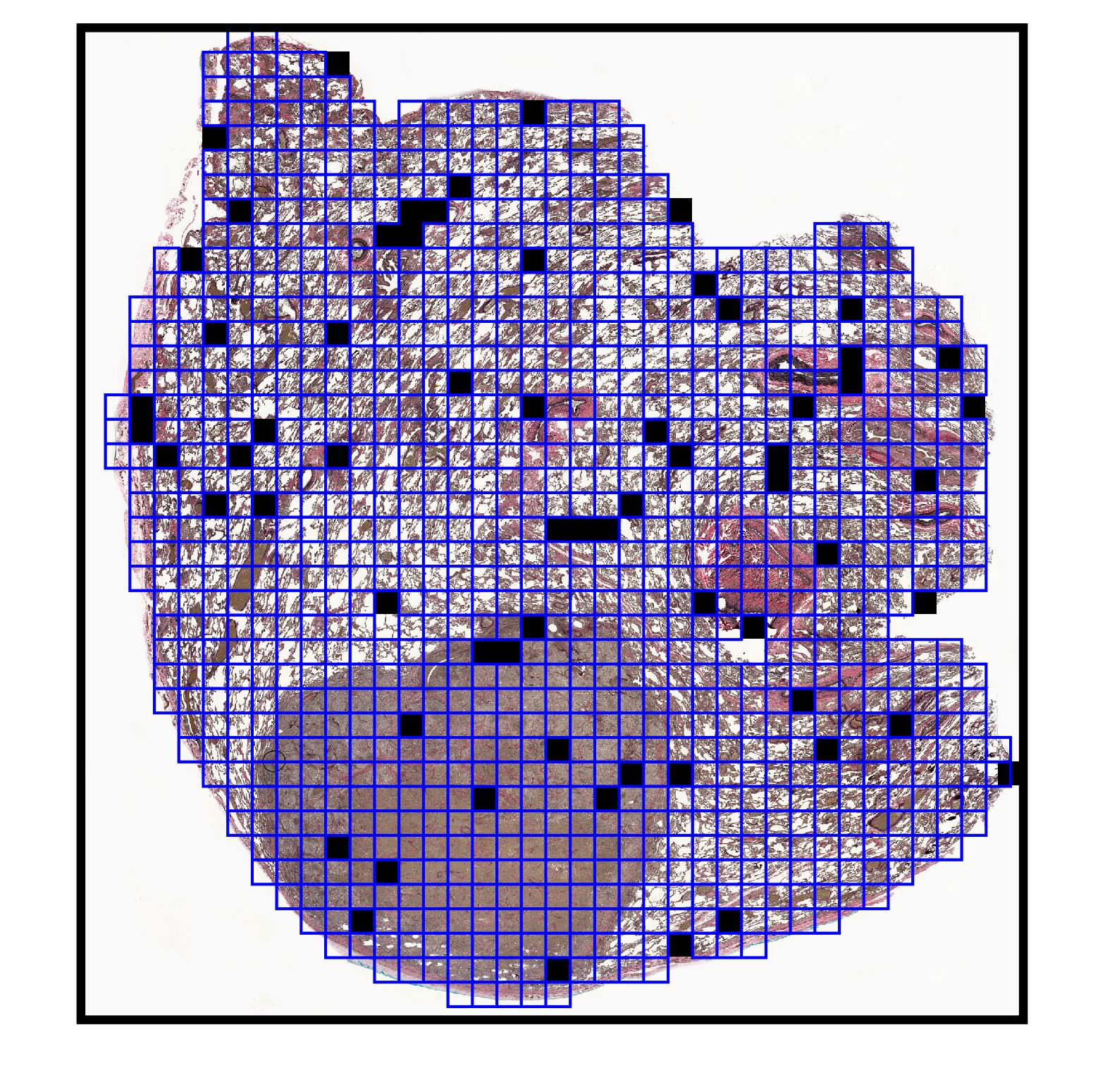}
\includegraphics[height=3cm, width=3.5cm]{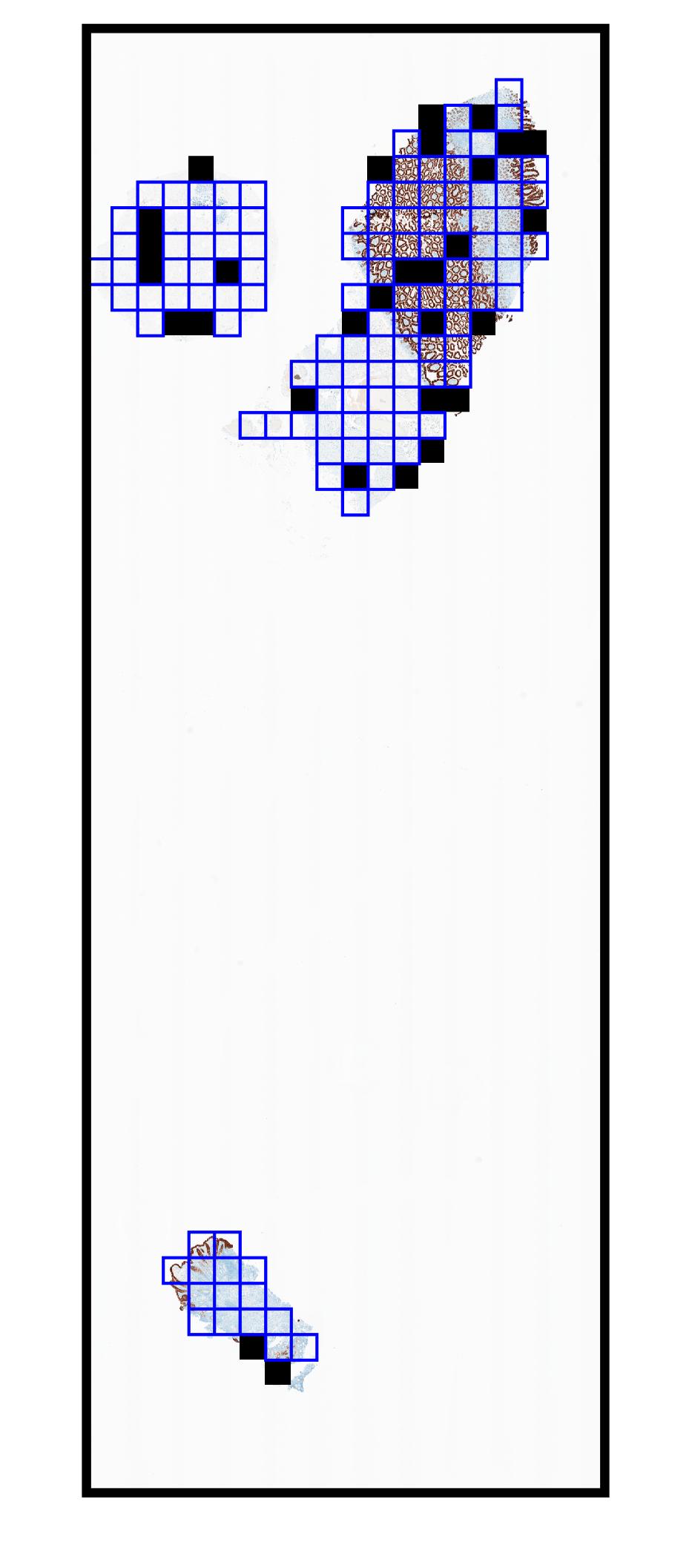}
\end{center}
\caption{Four examples of our segmentation results and patch extraction: the first row is the thumbnail of the slide, the outcome of the $5-$class K-means marked in red and the outcome of the $5-$class GMM marked in black. The locations of test patches are marked with black squares.}
\label{fig:segmentation}
\end{figure*}

From Table \ref{tab1}, we conclude that the size of the training dataset extracted based on the GMM is larger than K-means algorithm. 
The segmentation outputs of four WSIs obtained by two techniques and extracted patches with at most 50\% background are shown in Fig. \ref{fig:segmentation}. As Fig. \ref{fig:segmentation} shows, cluster assignment is much more flexible in GMM than in K-means. The latter loses many patches containing fat tissue by mistaking them with background pixels.

For the image search, one can compare the corresponding feature vectors of two images using a distance metric to measure the similarity. For three selected networks, we have used similarity metrics to find the most similar image in the training dataset for each query image in the test dataset based on the minimum distance between the deep features of the query and training images. In this study, We have utilized city block distance to measure the similarity between two feature vectors.

To evaluate the different strategies, we utilize different accuracy measures  \cite{babaie2017classification,kieffer2017convolutional}. The test dataset consists of $n=1325$ patches that belong to 24 WSIs. Suppose that $\Gamma_{s_i}=\{P^j_{s_i} ; j=1,\dots,n_i\}$ is the set of all patches for slide $i$ in the test dataset in which  $n_i$ is the number of test patches for slide $s_i$ and $i=0,1,\dots,23$. The patch-to-scan and whole-scan accuracy are defined as
$$
\eta_p=\frac{\sum_{i=0}^{23} |T_{s_i} \cap\Gamma_{s_i}|}{n}, ~~\hbox{and}~~
\eta_w=\frac{1}{24}\sum_{i=0}^{23} \frac{|T_{s_i} \cap\Gamma_{s_i}|}{n_i},
$$ 
respectively, where set $T_{s_i}$ contains the corresponding patches of slide $s_i$ in the training dataset and $|A|$ denotes the cardinality of set $A$. In addition, the total accuracy is given by
$\eta_{tot}=\eta_p\times\eta_w,$ 
which can be used to take into account both patch-to-scan and whole-scan accuracy values.

\begin{table*}[htb]
\begin{center}
\caption{The accuracy (in \%) for each strategy and each training dataset. The highest accuracy values are highlighted in bold.}
\label{tab2}
\resizebox{\textwidth}{!}{
\begin{tabular}{lllccccccccccccccc}
\hline
\textbf{Measure} & \textbf{Network} & \multicolumn{7}{c}{\textbf{Without Overlap}}  && \multicolumn{7}{c} {\textbf{With Overlap}}\\
\cline{4-8}  \cline{12-16}
 & & \multicolumn{7}{c}{\textbf{Background Percentage}}  && \multicolumn{7}{c}{\textbf{Background Percentage}} \\
\cline{3-9}   \cline{11-17}
 & &  10 & 20 & 30 & 40 & 50 & 60 & 70  & & 10 & 20 & 30 & 40 & 50 & 60 & 70\\
\hline
$\eta_p$     & Dense    & 76.53$^\dag$  & 81.36 & 85.51 & 87.25 & 90.11 & 93.06 & 94.19 &&  78.49 & 83.09 & 86.42 & 88.53 & 91.17 & 92.98 &  94.04 \\
             & Dense    & 83.85$^\ddag$ & 87.85 & 92.30 & 93.43 & 94.34 & 95.62 &\textbf{95.92} && 82.49 & 89.89 & 92.00 & 93.28 & 94.64 &   94.94 & 95.25 \\
             & IncV3    & 76.53 & 80.53 & 83.55 & 85.96 & 87.77 & 89.96 & 91.09  &&  78.11 & 81.96 & 84.23 & 87.02 & 89.36 & 90.72 & 91.47  \\                 
             & IncV3    & 81.13 & 86.19 & 89.58 & 90.94 & 91.85 & 92.08 & 92.45  &&  82.26 & 87.62 & 89.58 & 91.40 & 92.08 & 92.15 & 92.08  \\                  
             & VGG16    & 71.02 & 77.58 & 80.15 & 83.17 & 85.66 & 88.83 & 90.34  &&  74.57 & 80.75 & 83.25 & 85.51 & 87.40 & 88.91 & 89.58  \\  \vspace{1mm}    
             & VGG16    & 79.77 & 83.77 & 87.32 & 89.58 & 90.64 & 91.55 & 92.00  &&  81.21 & 85.74 & 88.68 & 89.36 & 90.04 & 90.49 & 90.64  \\            
$\eta_w$     & Dense    &  74.67$^\dag$ & 79.54 & 83.62 & 85.42 & 88.72 & 91.37 & 92.91 &&  77.18 & 81.74 & 84.90 & 86.81 & 89.62 & 91.60 & 92.79   \\                          & Dense    & 83.85$^\ddag$ & 87.67 & 91.76 & 92.93 & 93.83 & 95.13 & \textbf{95.51} &&  81.91 & 88.56 & 90.91 & 92.20 & 93.42 & 93.92 & 94.30   \\      
             & IncV3    & 74.04 & 78.40 & 81.57 & 83.66 & 85.65 & 87.95 & 89.28   &&  77.31 & 80.96 & 82.82 & 85.80 & 88.03 & 89.36 & 90.23   \\                
             & IncV3    & 80.51 & 85.12 & 88.40 & 89.61 & 90.68 & 90.99 & 91.40   &&  81.30 & 86.43 & 88.53 & 90.33 & 91.04 & 91.03 & 91.05    \\                
             & VGG16    & 68.89 & 75.50 & 77.87 & 80.97 & 83.99 & 86.92 & 88.51   &&  73.73 & 79.45 & 81.86 & 83.90 & 86.26 & 87.85 & 88.70   \\  \vspace{1mm}   
             & VGG16    & 78.75 & 82.56 & 86.11 & 88.38 & 89.53 & 90.42 & 90.86   &&  79.49 & 84.04 & 87.16 & 87.73 & 88.42 & 88.97 & 89.15    \\          
$\eta_{tot}$ & Dense    & 57.14$^\dag$  & 64.71 & 71.50 & 74.53 & 79.95 & 85.02 & 87.51 &&  60.58 & 67.92 & 73.37 & 76.85 & 81.70 & 85.17 & 87.26 \\                            & Dense    & 70.31$^\ddag$ & 77.01 & 84.69 & 86.82 & 88.52 & 90.96 & \textbf{91.62} && 67.57 & 79.61 & 83.6 & 86.01 & 88.42 & 89.17 & 89.82 \\   
             & IncV3    & 56.66 & 63.14 & 68.15 & 71.91 & 75.18 & 79.12 & 81.33  && 60.39 & 66.36 & 69.75 & 74.66 & 78.66 & 81.06 & 82.53 \\                 
             & IncV3    & 65.32 & 73.37 & 79.19 & 81.50 & 83.29 & 83.78 & 84.50  && 66.88 & 75.74 & 79.31 & 82.56 & 83.83 & 83.89 & 83.84 \\   
             & VGG16    & 48.92 & 58.57 & 62.41 & 67.34 & 71.95 & 77.21 & 79.96  && 54.98 & 64.16 & 68.15 & 71.74 & 75.38 & 78.10 & 79.46   \\   
             & VGG16    & 62.82 & 69.16 & 75.19 & 79.17 & 81.15 & 82.78 & 83.59  && 64.55 & 72.05 & 77.29 & 78.39 & 79.61 & 80.51 & 80.81  \\                           
\hline
\multicolumn{15}{l}{$^\dag$ and $^{\ddag}$ stand for the K-means and GMM, respectively.}
\end{tabular}
}
\end{center}
\end{table*}

For comparison of all strategies and selecting the best training dataset, the accuracy for each strategy are summarized in Table \ref{tab2}.
As shown in Table \ref{tab2}, the \textit{DenseNet} yields the best results. Accuracy values obtained with the \textit{VGG16} network are rather low. In addition, the accuracy of extracted training dataset without any overlap is higher than the training dataset with 20\% overlap. Also, the GMM strategy is superior to K-means segmentation. 
From the results shown in  Table \ref{tab2}, it is apparent that the highest accuracy values belong to training data with 22,591 patches which were extracted using GMM without any overlap and at most 70\% background. We considered this set as the training set for the \emph{Kimia Path24C}  dataset and make it publicly available.


Concerning the file size in archives of histopathology images, low storage demand and high search speed are crucial for feasible CBIR systems. We used the concept of \textbf{deep barcodes} \cite{barcode1, barcode2} to compare the performance of deep features with deep barcodes. As Table \ref{tabBarcodes} shows deep barcodes do suffer from a slight drop in accuracy value. However, the practical benefits of using binary information may over-weigh the small accuracy loss.

\begin{table}[htb]
\begin{center}
\caption{Comparison of deep features and deep barcodes.}
\label{tabBarcodes}
\resizebox{\columnwidth}{!}{
\begin{tabular}{lccc}
\hline
\textbf{Measure} & \textbf{Network} \\
\cline{2-4} 
 & \textit{VGG16} & \textit{InceptionV3} & \textit{DenseNet-121} \\
\hline
$\eta_p$      &   92.00$^\dag$   &  92.45  &  \bf{95.92} \\ \vspace{2mm}
$\eta_p$      &   86.57$^\ddag$  &  91.17  &  93.43  \\
$\eta_w$      &   90.86  &  91.40  &  \bf{95.51}  \\   \vspace{2mm}
$\eta_w$      &   85.21  &  89.75  & 92.37   \\
$\eta_{tot}$  &   83.59  &  84.50  & \bf{91.62}  \\       
$\eta_{tot}$  &   73.76  &  81.83  & 86.30  \\
\hline
\multicolumn{4}{l}{$^\dag$ and $^{\ddag}$ stand for deep features and barcodes, respectively.}
\end{tabular}
}
\end{center}
\end{table}



\begin{figure*}[hb]
\begin{center}
\includegraphics[height=3.5cm, width=3.5cm]{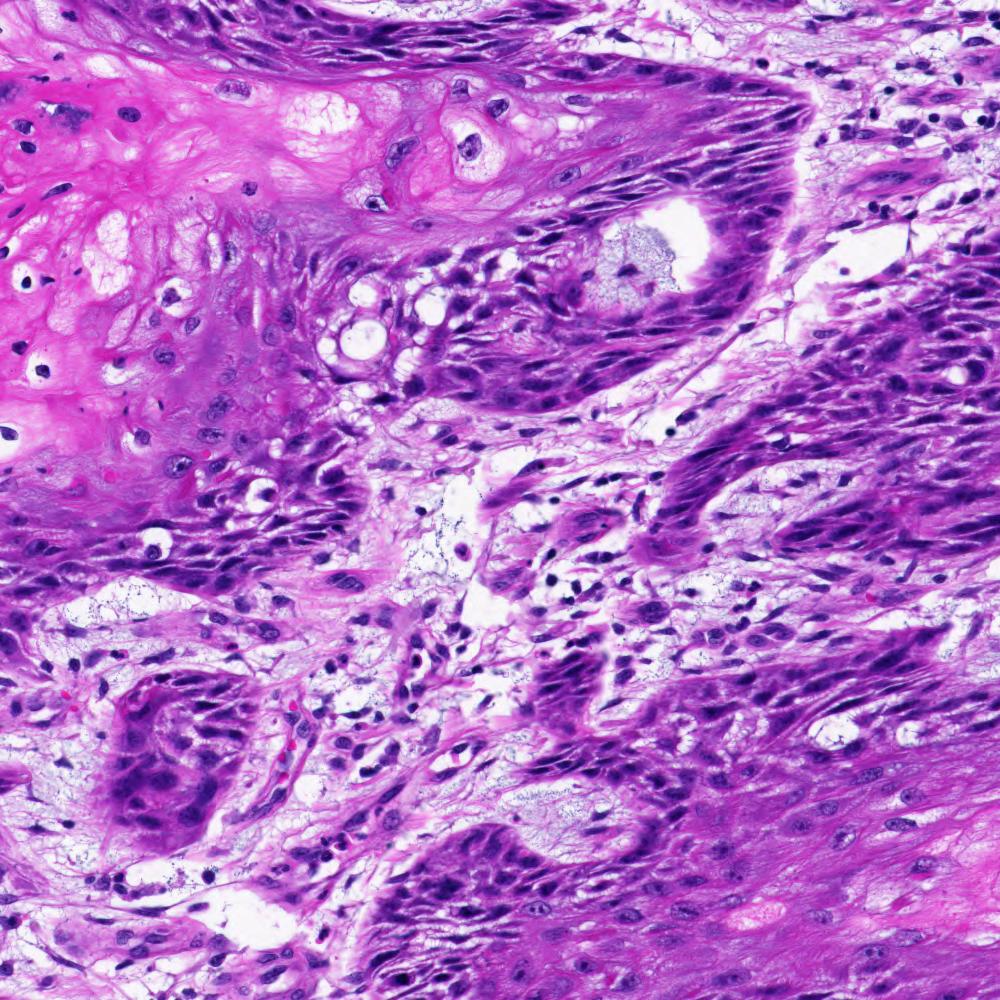}
\includegraphics[height=3.5cm, width=3.5cm]{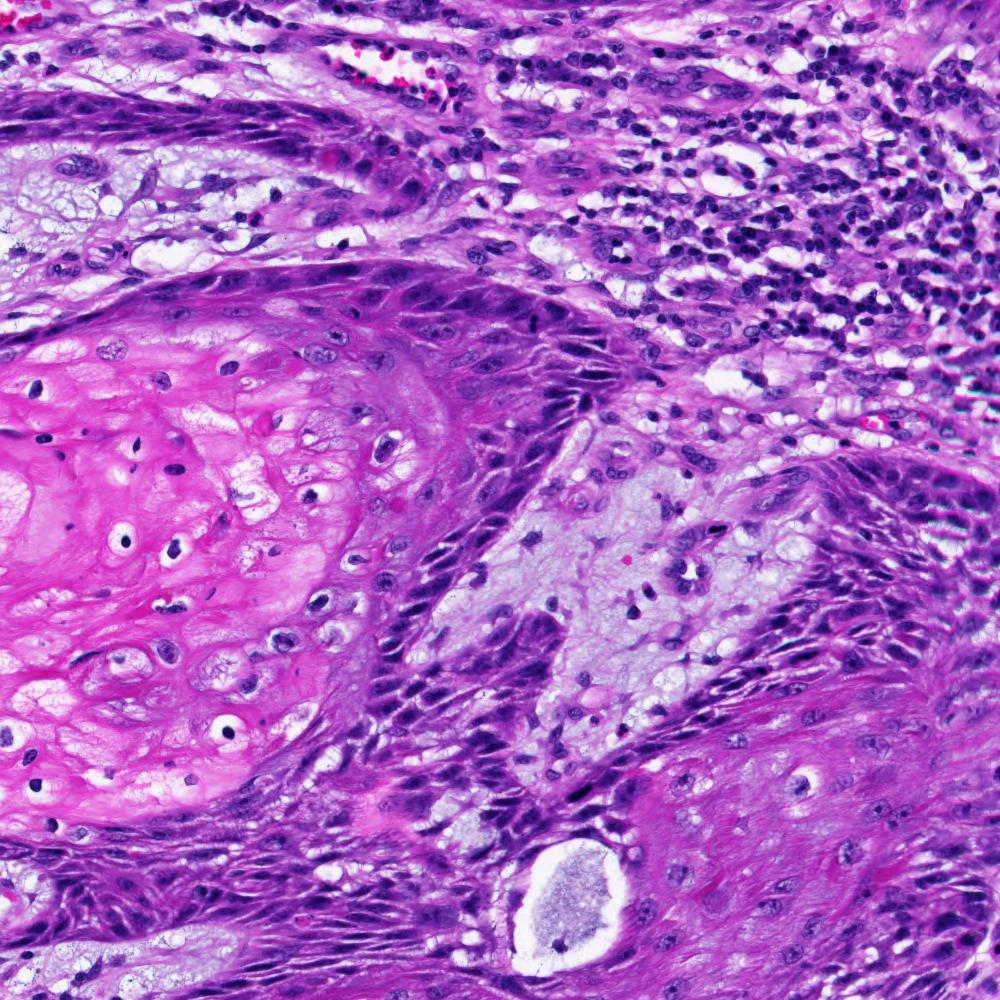}
\includegraphics[height=3.5cm, width=3.5cm]{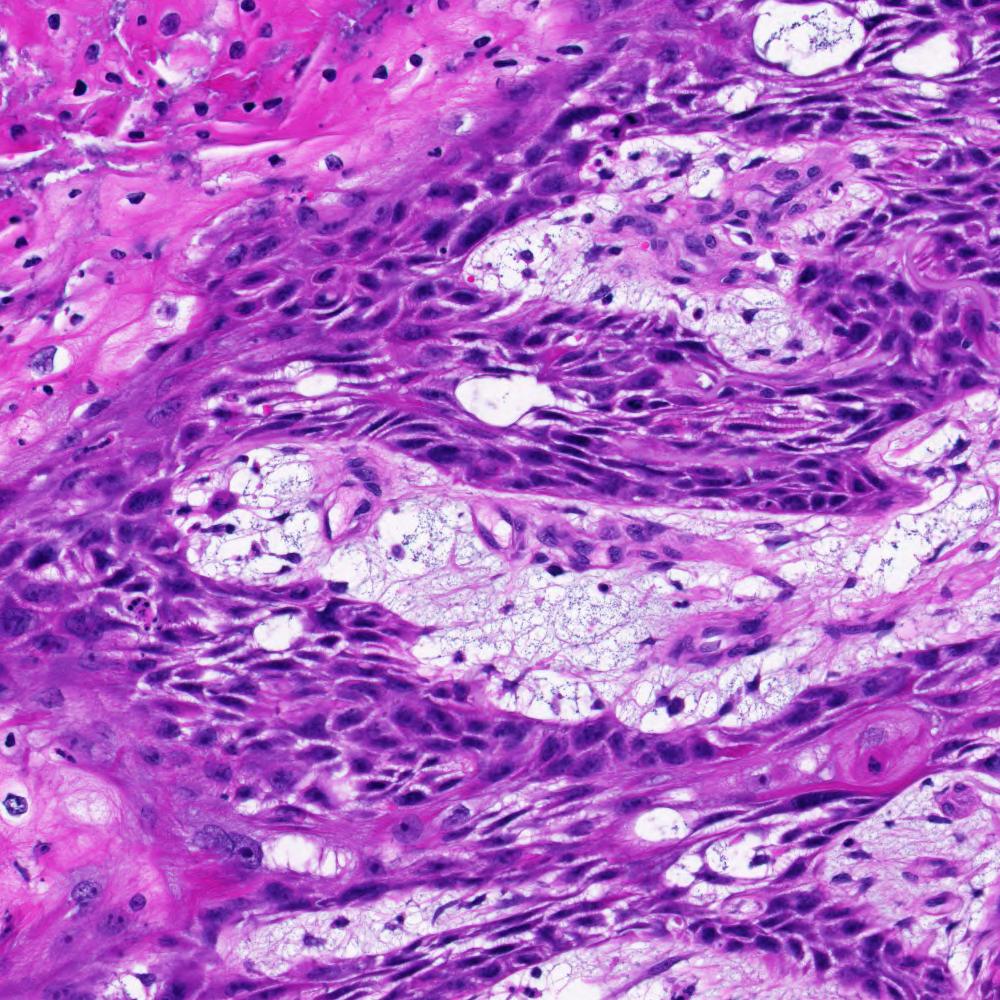}
\includegraphics[height=3.5cm, width=3.5cm]{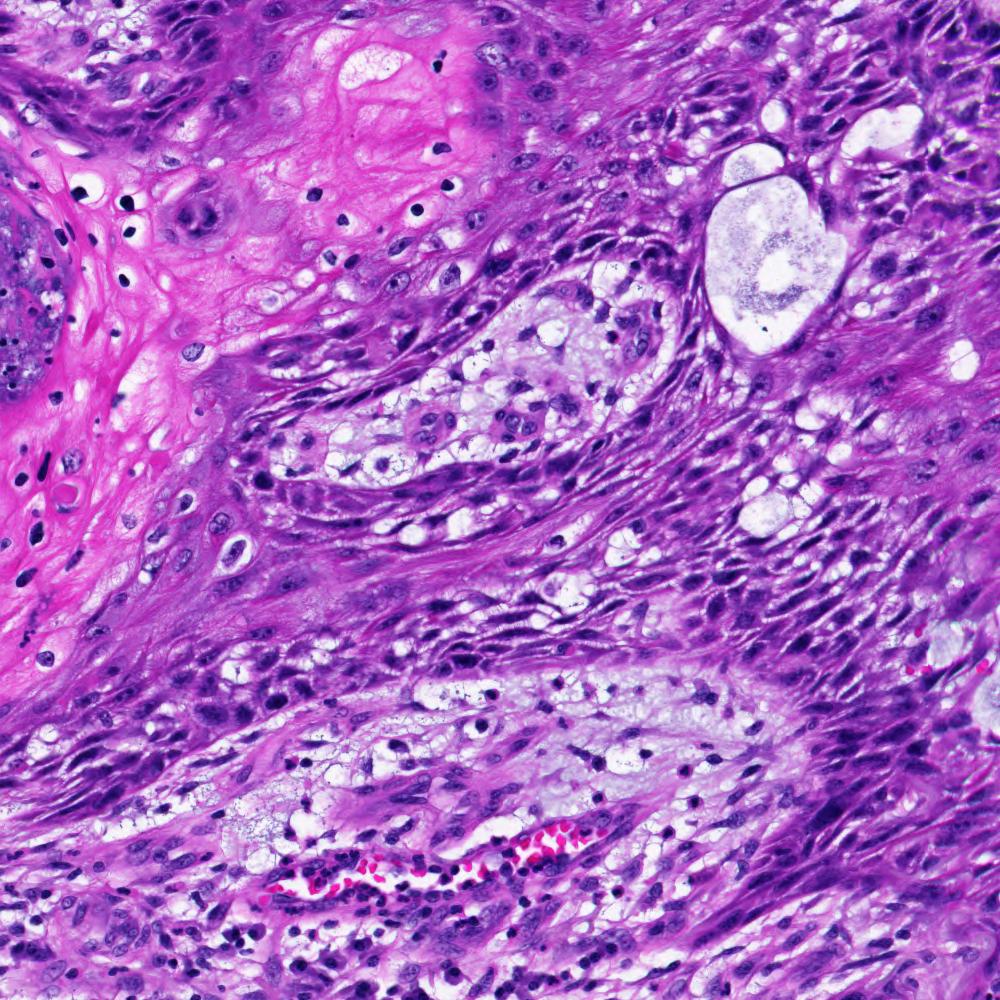}
\includegraphics[height=3.5cm, width=3.5cm]{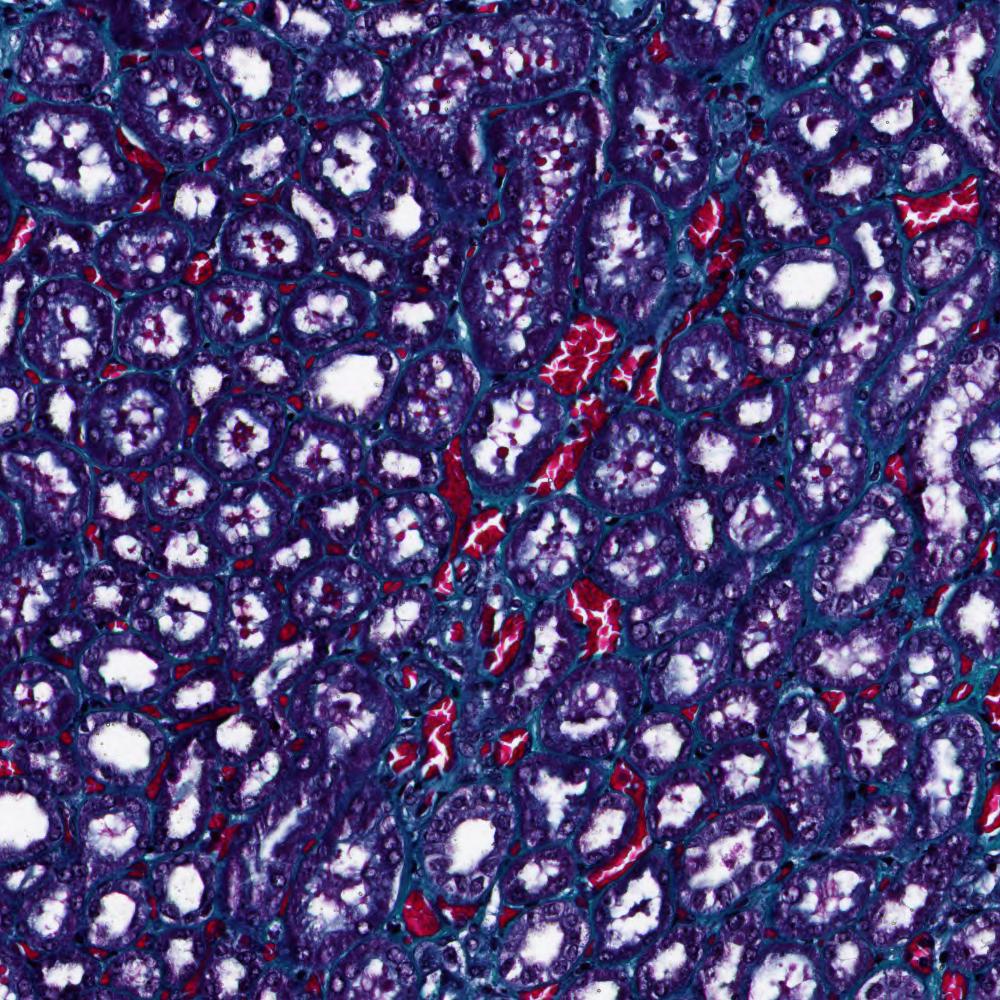}
\includegraphics[height=3.5cm, width=3.5cm]{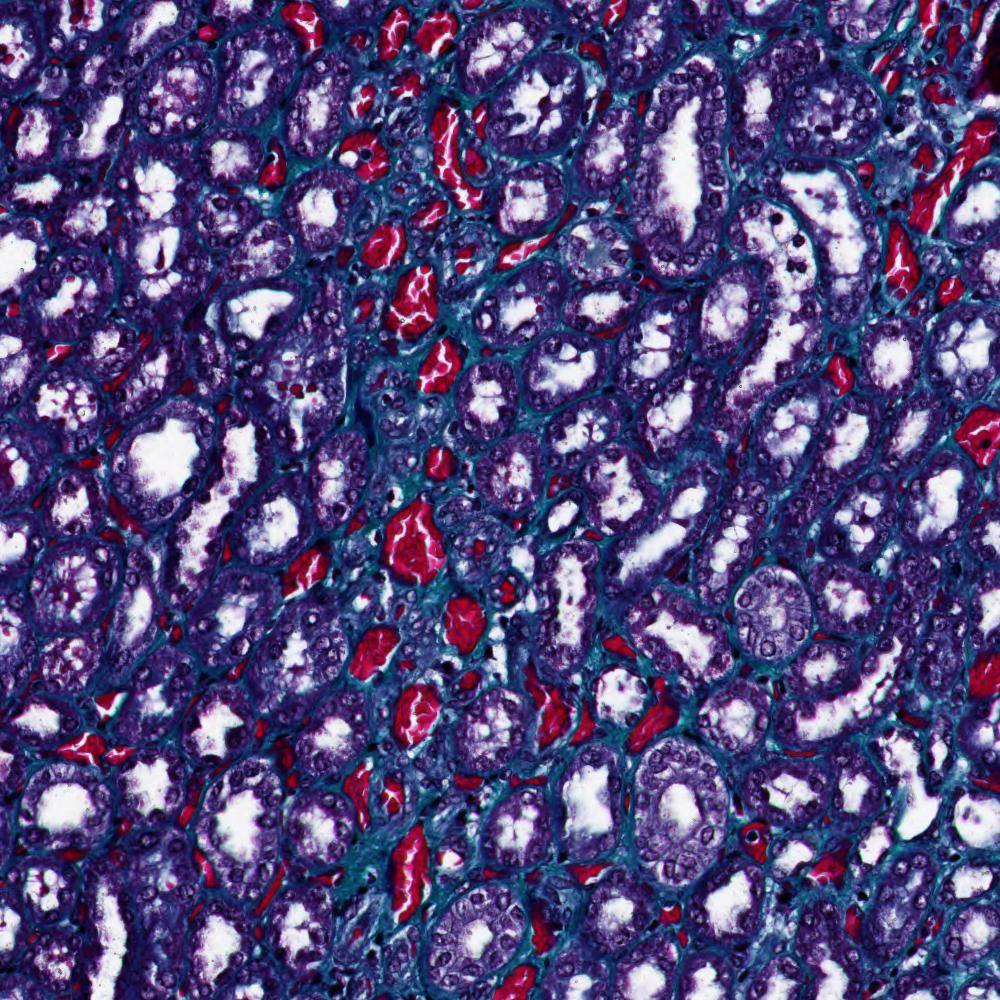}
\includegraphics[height=3.5cm, width=3.5cm]{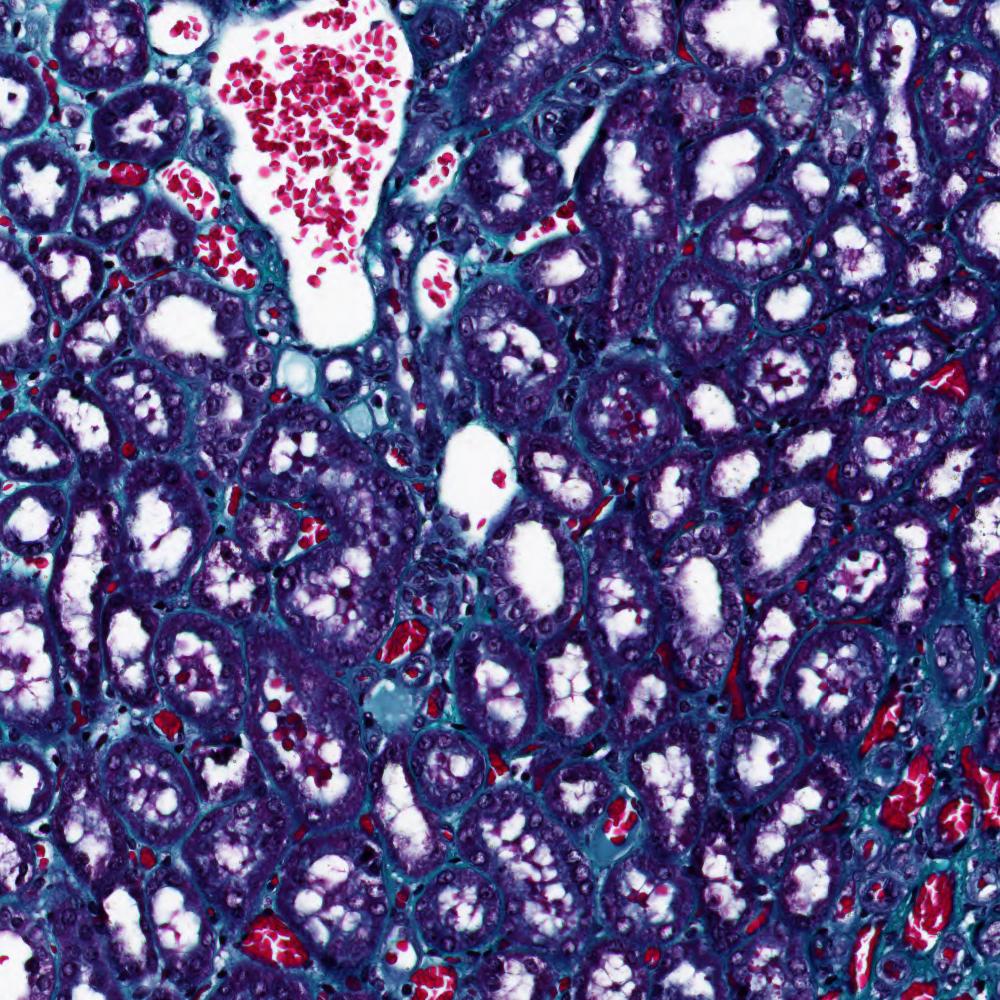}
\includegraphics[height=3.5cm, width=3.5cm]{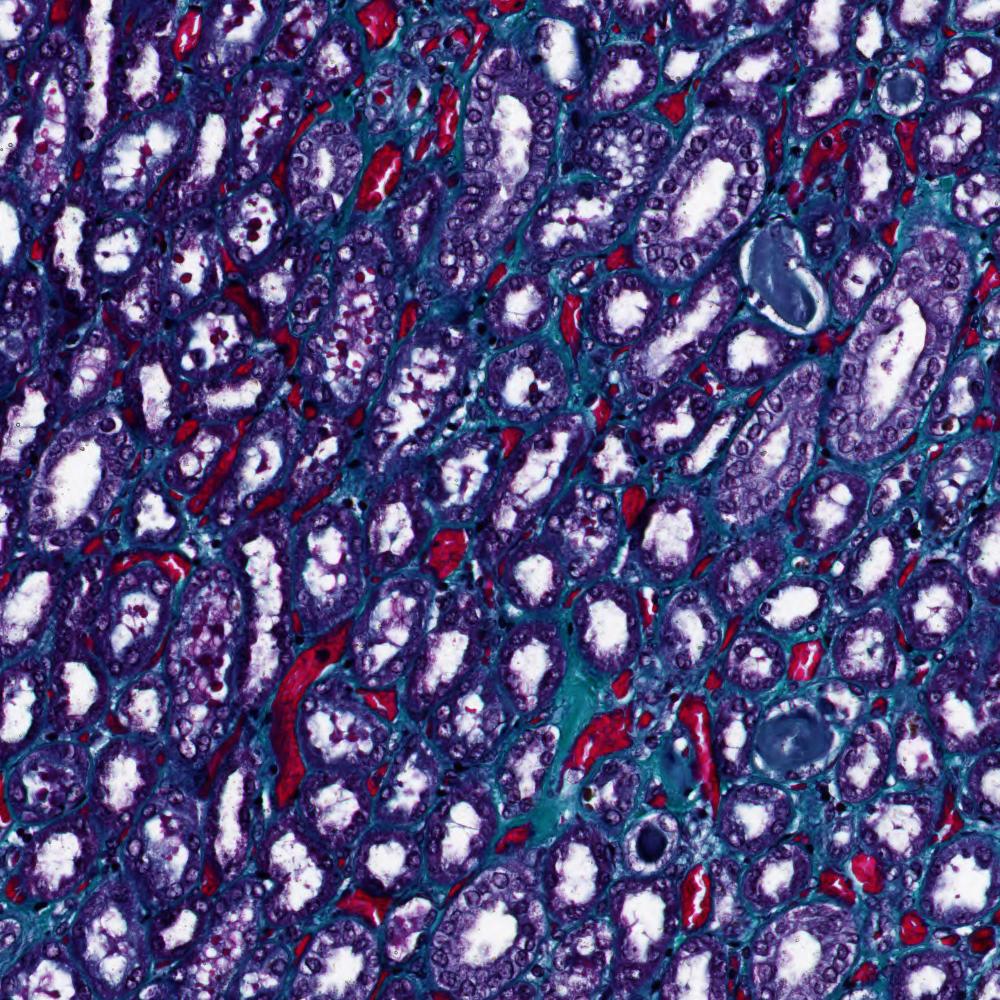}
\includegraphics[height=3.5cm, width=3.5cm]{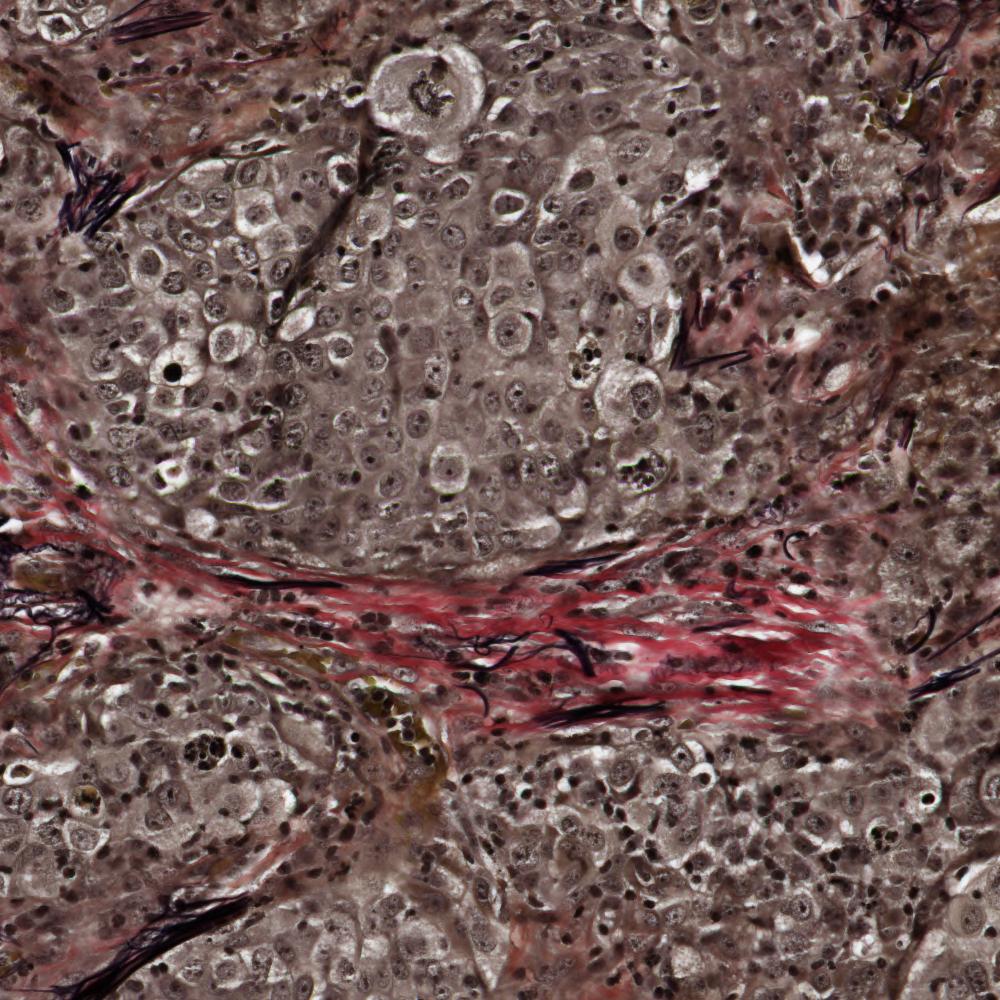}
\includegraphics[height=3.5cm, width=3.5cm]{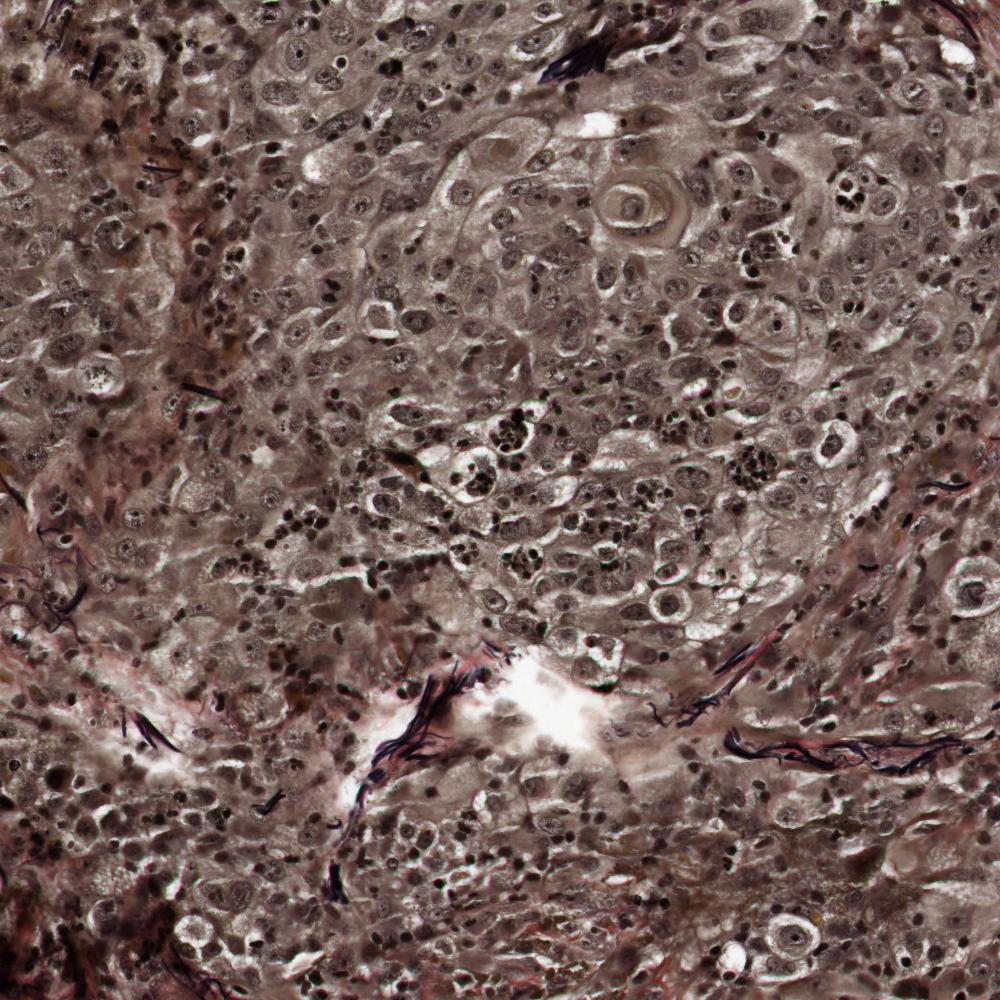}
\includegraphics[height=3.5cm, width=3.5cm]{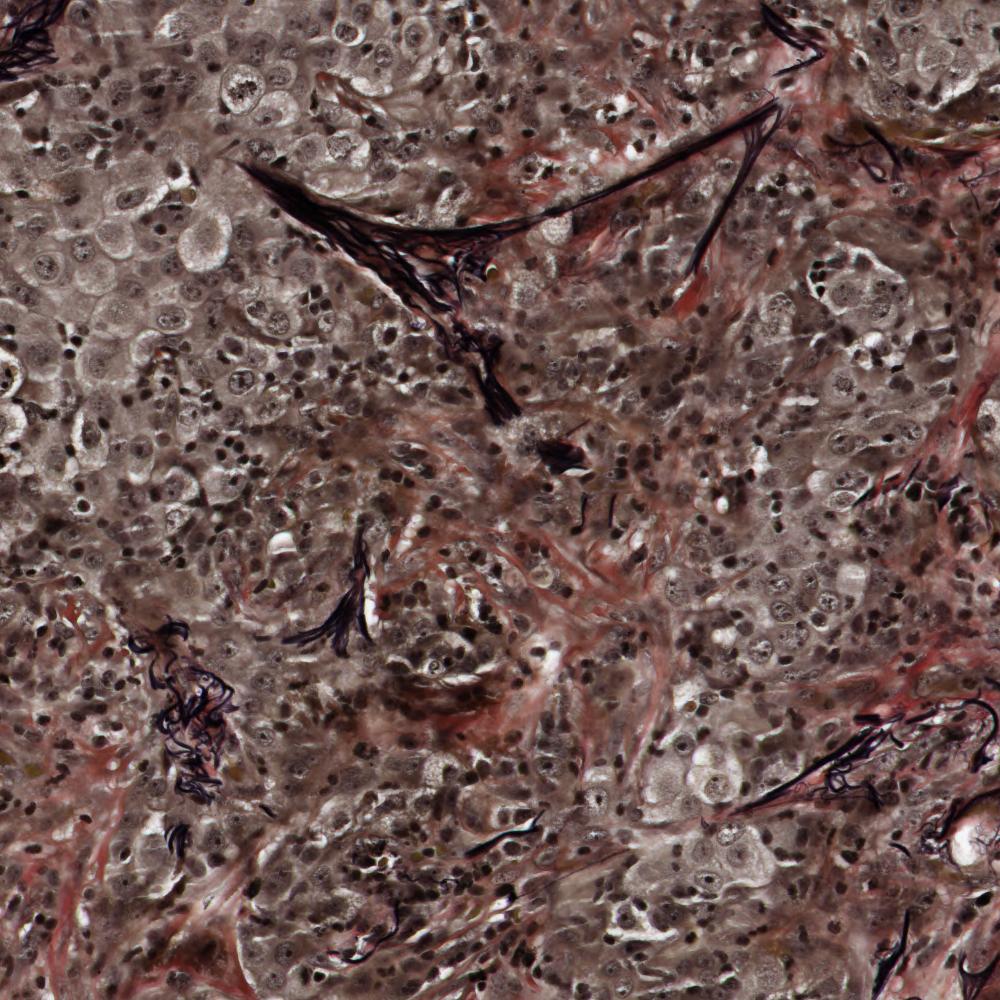}
\includegraphics[height=3.5cm, width=3.5cm]{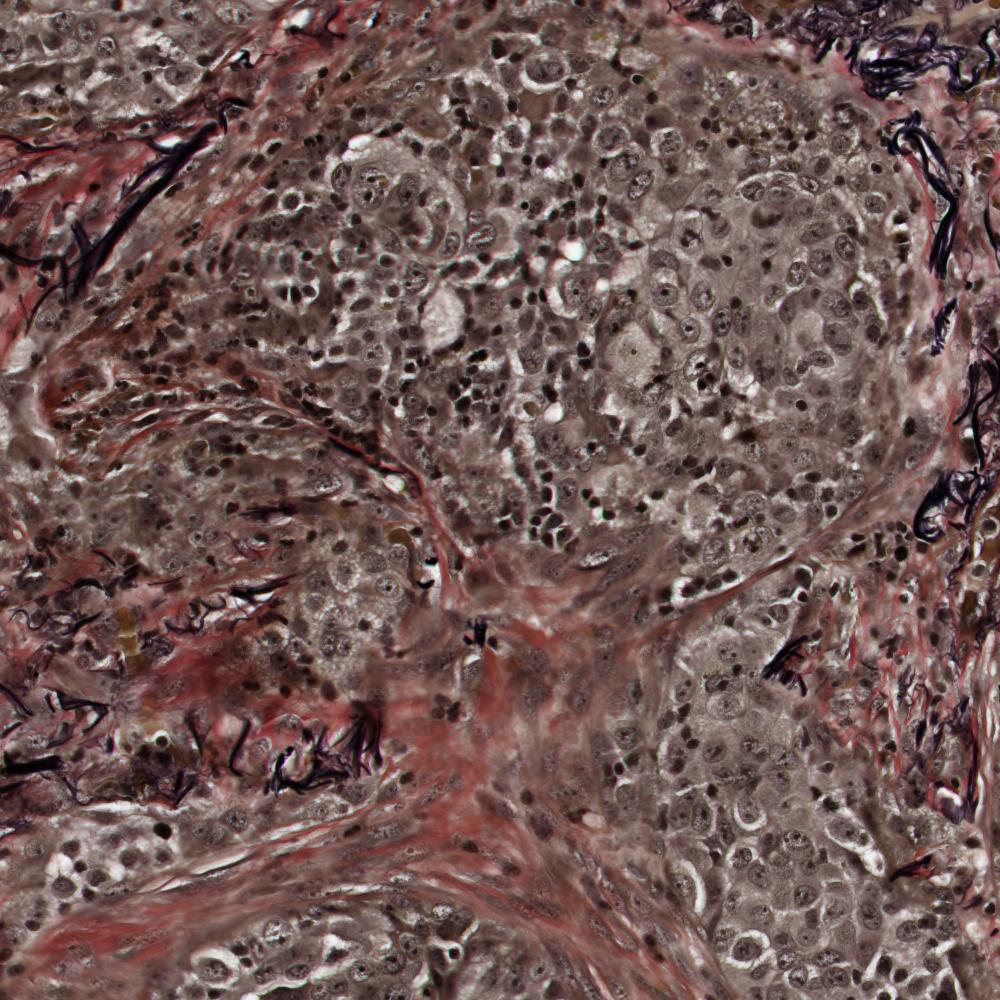}
\includegraphics[height=3.5cm, width=3.5cm]{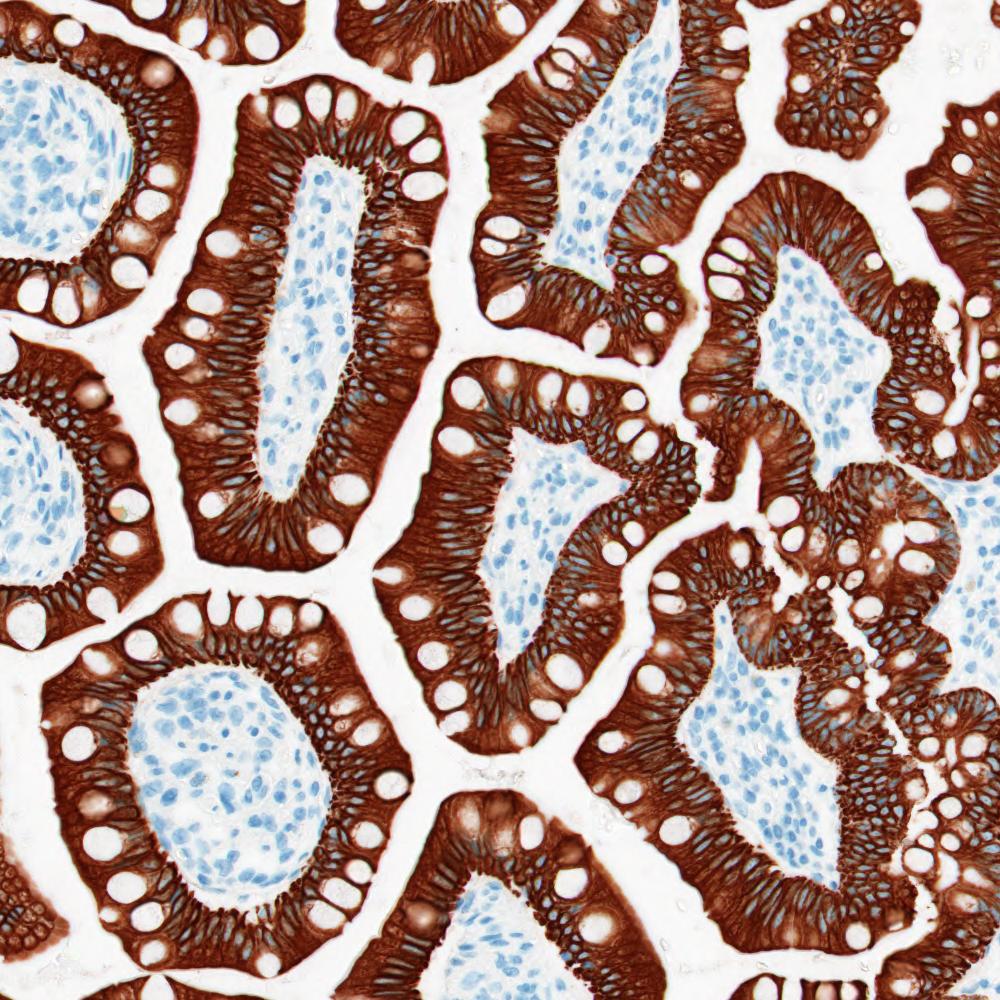}
\includegraphics[height=3.5cm, width=3.5cm]{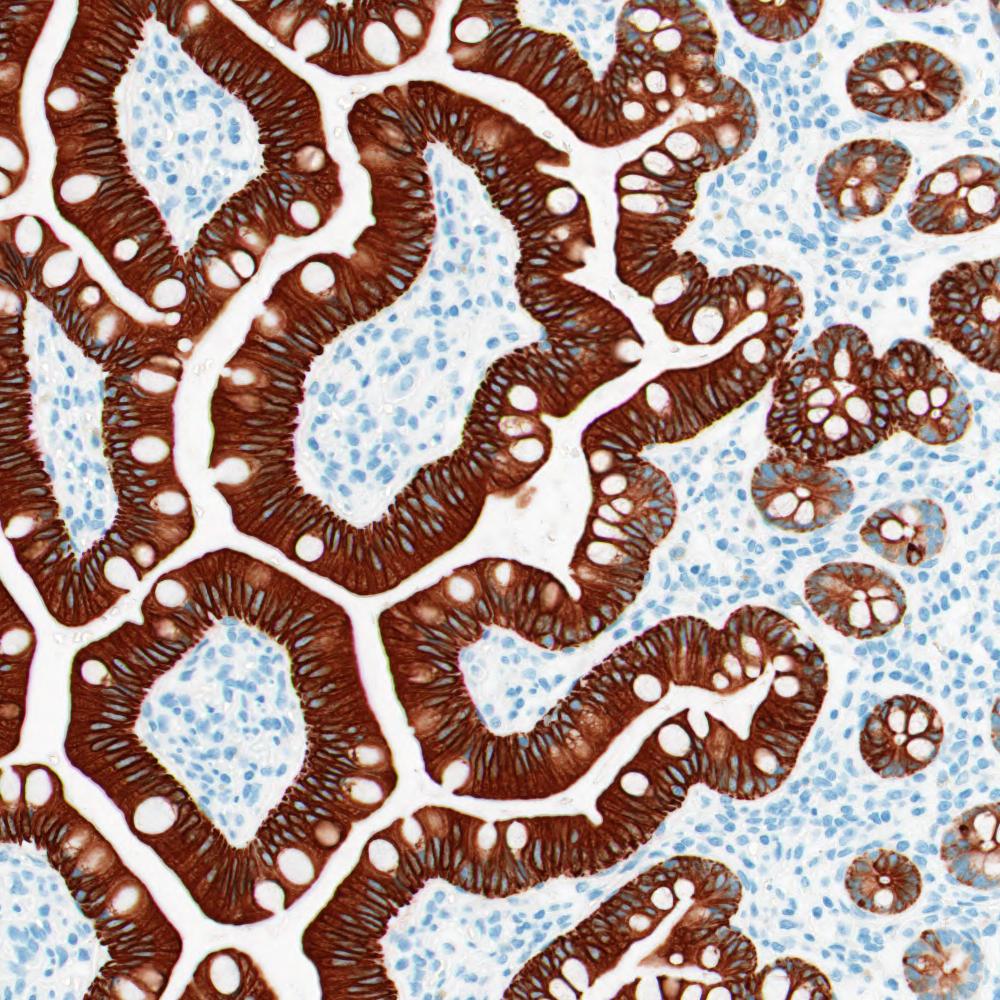}
\includegraphics[height=3.5cm, width=3.5cm]{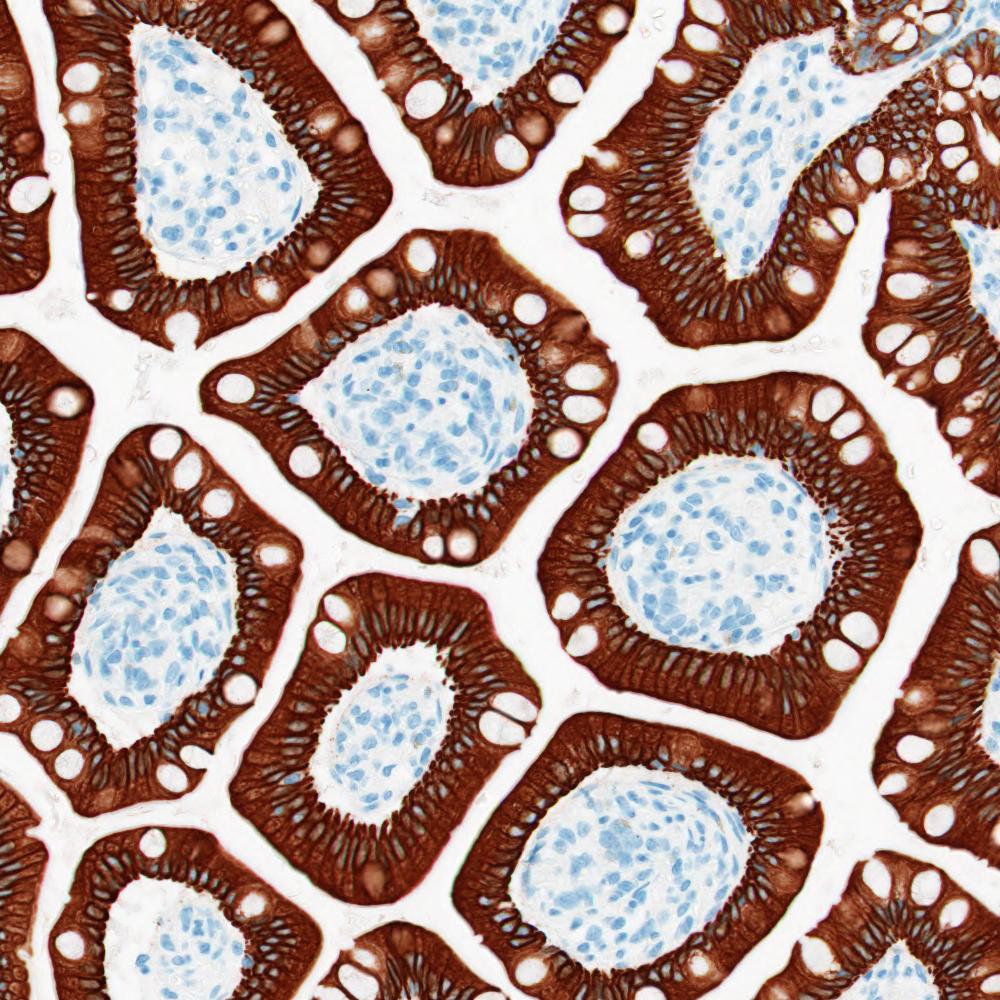}
\includegraphics[height=3.5cm, width=3.5cm]{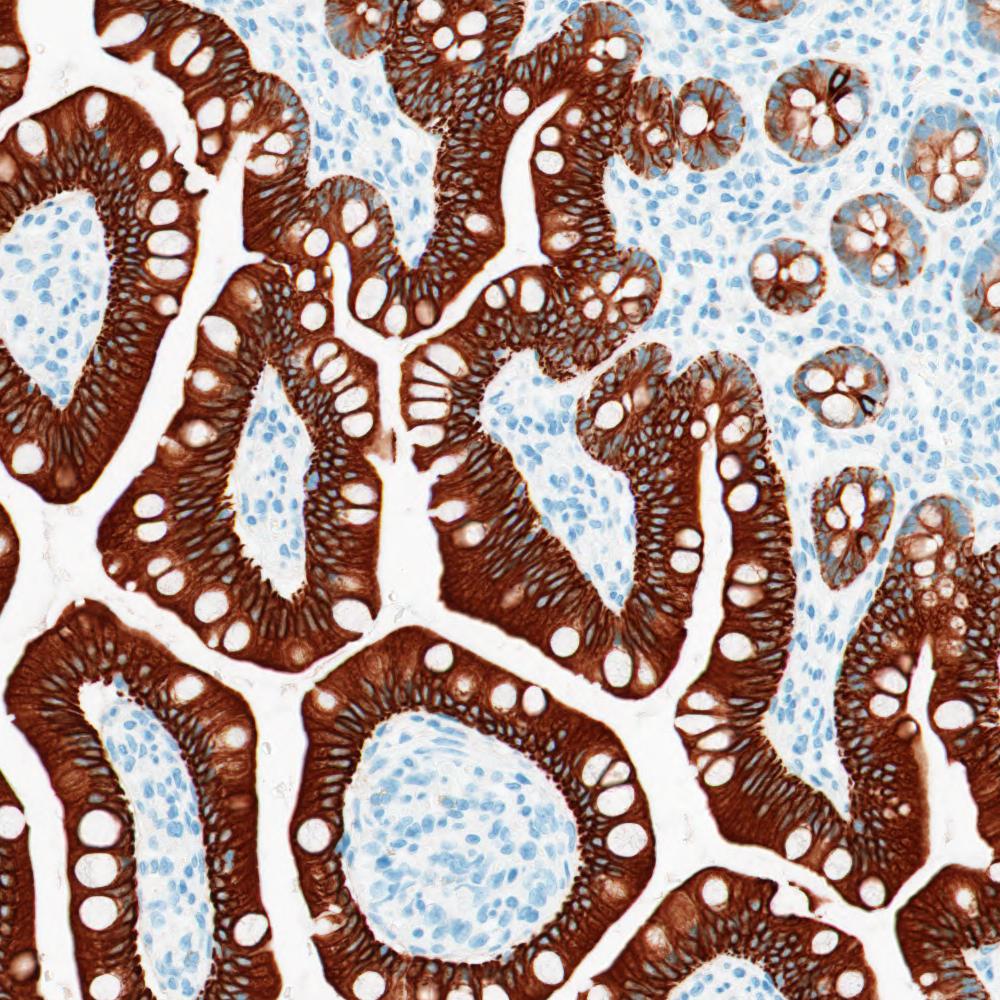}
\includegraphics[height=3.5cm, width=3.5cm]{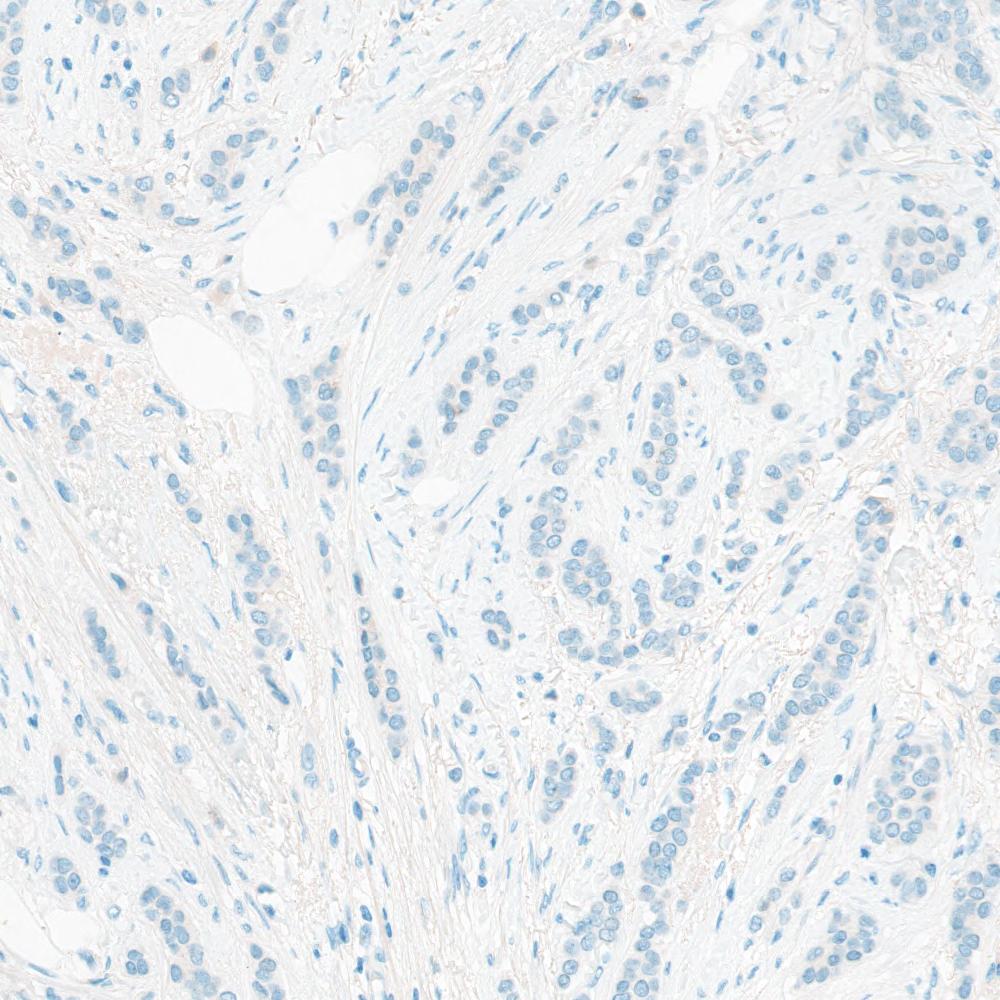}
\includegraphics[height=3.5cm, width=3.5cm]{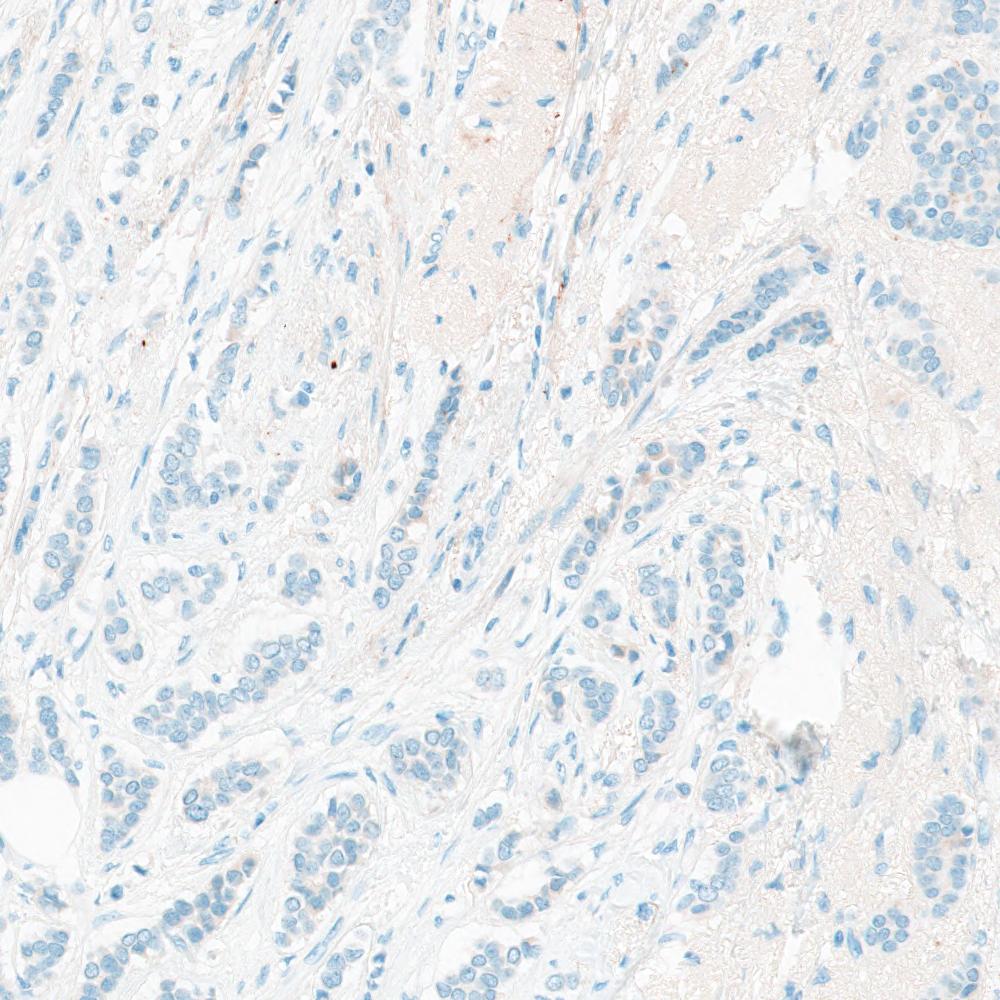}
\includegraphics[height=3.5cm, width=3.5cm]{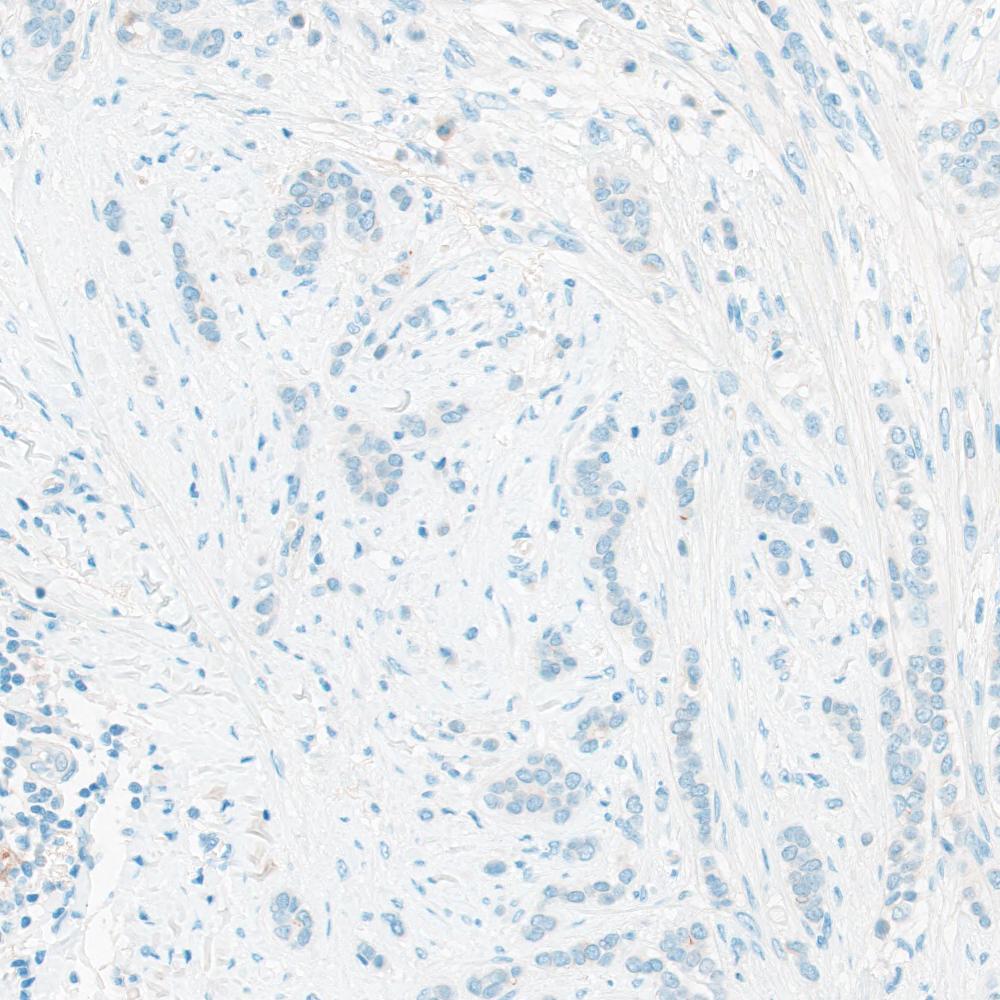}
\includegraphics[height=3.5cm, width=3.5cm]{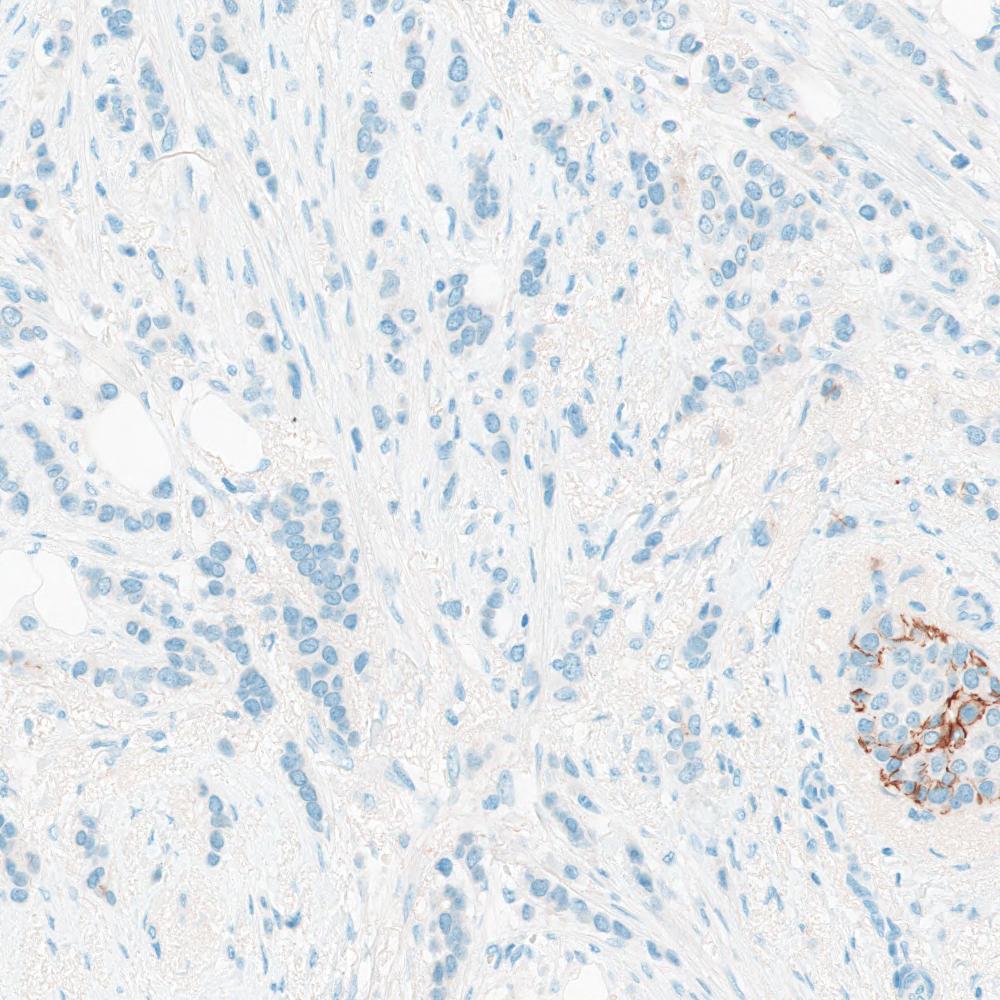}
\end{center}
\caption{Five examples of image retrieval results. The first column contains the query patches, and the other columns are top-3 retrieved images.}
\label{fig:retri}
\end{figure*}

Fig. \ref{fig:retri} shows some retrieval examples of our image retrieval experiments based on the selected training set and the \textit{DenseNet} model as a feature extractor. The top three relevant images are listed for each query image. The outstanding patch retrieval accuracy has applications in digital pathology such as floater detection when we are interested to find the same tissue pattern originating from the same patient. 

\subsection{Summary and Conclusions}
In this paper, we have presented a study on determining a suitable training dataset of RGB patches necessary to achieve high accuracy for the Kimia Path24 dataset. For each of the 24 slides, patches were extracted with several different values for the foreground/background and the patch overlap ratios. Next, we selected a total of 22,591 RGB patches as the training dataset based on the accuracy of image retrieval. Although the number of extracted patches is less than the grayscale training dataset proposed by \cite{babaie2017classification} but the accuracy of image search has been increased. The highest accuracy of image retrieval using proposed training data is 95.92\% while the highest accuracy of grayscale training dataset based on the Local Binary Patterns \cite{babaie2017classification} and Fine-tuning \textit{InceptinV3} \cite{kieffer2017convolutional} reached 66.11\% and 74.87\%, respectively. The new dataset, called \emph{Kimia Path24C}, is publicly available at {\protect\url{http://kimia.uwaterloo.ca/}}.

%

{\small
\bibliographystyle{ieee}
\bibliography{ref}
}

\end{document}